\documentclass[manuscript]{acmart}

\usepackage{caption}
\usepackage{subcaption}
\usepackage{soul}
\usepackage{graphicx}
\usepackage{enumitem}
\usepackage{ulem}
\usepackage{makecell}

\newcommand{\note}[1]{\textcolor{red}{#1}}
\newcommand{\subsubsub}[1]{\textbf{#1}}

\newcommand{\cut}[1]{}
\newcommand{\add}[1]{\textcolor{black}{#1}}
\newcommand{\jkk}[1]{\textcolor{green}{JKK: #1}}

\AtBeginDocument{%
  \providecommand\BibTeX{{%
    \normalfont B\kern-0.5em{\scshape i\kern-0.25em b}\kern-0.8em\TeX}}}

\setcopyright{none}
\begin{document}

\title{\add{Supporting Sensemaking of Large Language Model Outputs at Scale}}

\author{Katy Ilonka Gero}
\email{katy@g.harvard.edu}
\affiliation{%
  \institution{Harvard University}
  \country{USA}
}

\author{Chelse Swoopes}
\email{cswoopes@g.harvard.edu}
\affiliation{%
  \institution{Harvard University}
  \country{USA}
}

\author{Ziwei Gu}
\email{ziweigu@g.harvard.edu}
\affiliation{%
  \institution{Harvard University}
  \country{USA}
}

\author{Jonathan K. Kummerfeld}
\email{jonathan.kummerfeld@sydney.edu.au}
\affiliation{%
  \institution{University of Sydney}
  \country{Australia}
}

\author{Elena L. Glassman}
\email{glassman@seas.harvard.edu}
\affiliation{%
  \institution{Harvard University}
  \country{USA}
}

\begin{abstract}
\cut{Many large language model (LLM) interfaces support viewing just one or two outputs at a time, making it difficult to characterize model capabilities or find the best output for a given task.}
\add{Large language models (LLMs) are capable of generating multiple responses to a single prompt, yet little effort has been expended to help end-users or system designers make use of this capability.}
In this paper, we explore how to present many LLM responses at once. We design five features, which include both pre-existing and novel methods for computing similarities and differences across textual documents, as well as how to render their outputs. We report on a controlled user study (n=24) and \cut{two} \add{eight} case studies evaluating these features and how they support users in different tasks. We find that the features \cut{decreased perceived working memory load and supported users in creating and confirming hypotheses about model characteristics} \add{support a wide variety of sensemaking tasks and even make tasks previously considered to be too difficult by our participants   now tractable}. Finally, we present design guidelines to inform future explorations of new LLM interfaces.
\end{abstract}



\begin{teaserfigure}
    \centering
         \centering
         \includegraphics[width=\textwidth]{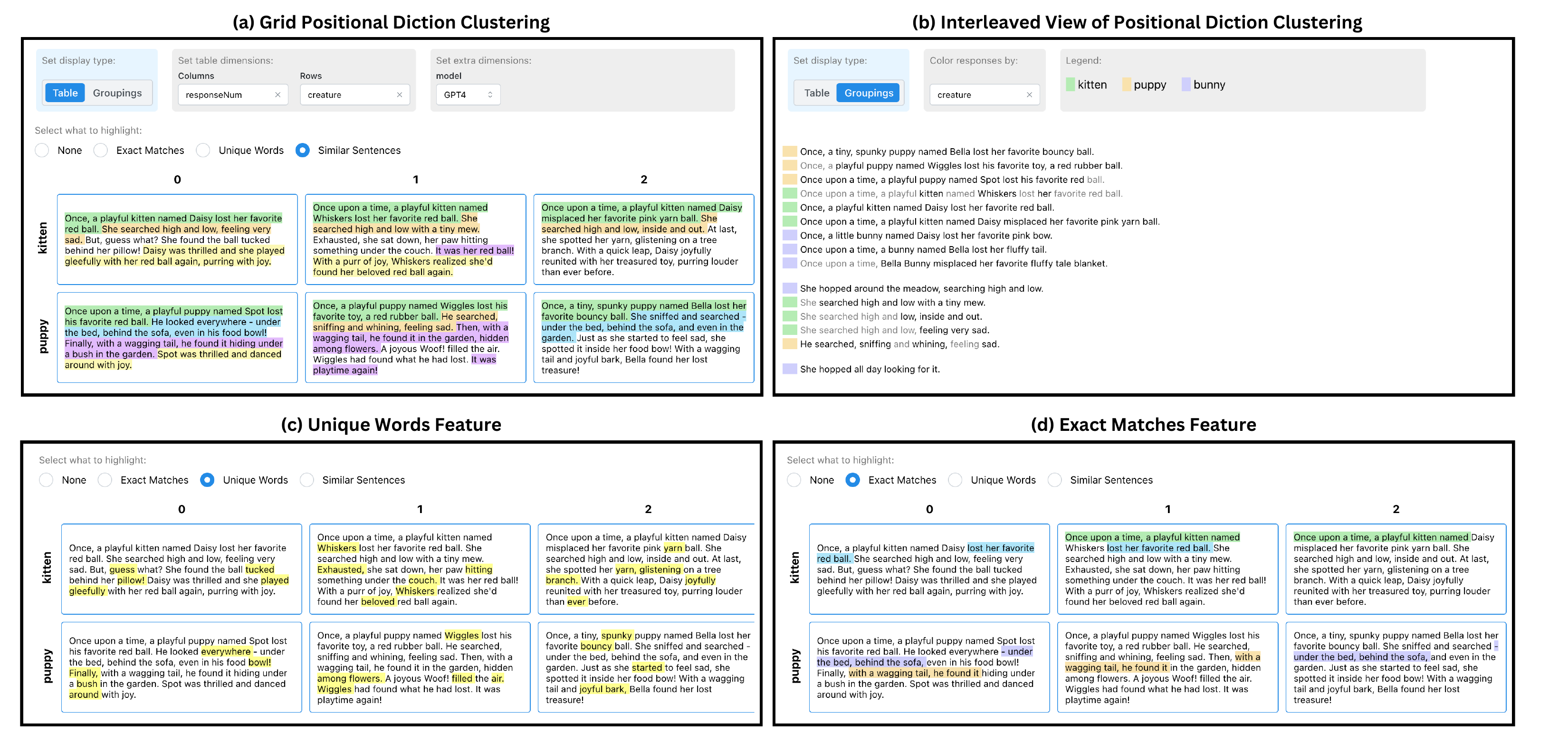}
	\caption{\add{Our exploratory interface instantiates five chosen combinations of text analysis (e.g., unique words, exact matches, and a novel algorithm we call Positional Diction Clustering) and renderings (within a grid with highlights or interleaved with grayed out redundancies), which can help users scale up the amount of LLM outputs they can reason about, e.g., for ideation, model comparison, or output selection. 
 These figures represent how four of the five different combinations render the top of a page rendering the entirety of a large collection of LLM outputs generated from the test prompt ``Write a short story for a five year old child about a \{creature\} that loses something and then finds it again.'' The fifth view tested, i.e., a grid layout without any visual additions based on text analysis, is not shown.}
	}
  \Description{Set of four screenshots of the system. Top left is the Grid View of Positional Diction Clustering, with a grid of six outputs where sentences are highlighted based on their positional and semantic similarity. There are five colors, representing five groups of sentences. Not every sentence is highlighted. Top right is the Interleaved View of Position Diction Clustering, where sentences are on individual lines and if a word in a sentence occurs in the same location in the previous sentence, that word is greyed out. To the left of the sentences is a colored box that indicates which prompt variation the sentence came from. Bottom left is the Unique Words Feature, where words more likely to occur in that cell than any other cells are highlighted. All words are highlighted in the same color. Bottom right is the Exact Matches Feature, where exact text segment matches are highlighted. There are four colors, indicated four sets of text segments that occurred in multiple responses.}
  \label{fig:teaser}
\end{teaserfigure}


\maketitle

\section{Introduction}

\add{Large language models are capable of generating multiple different responses to a single prompt. While several tools have recently been developed for structuring the generation of prompts and collecting responses~\cite{promptfoo, weightsandbiases, arawjo2023chainforge}, relatively little effort has been expended to help either end-users or designers of LLM-backed systems to reason about or make use of the variation seen in the multiple responses generated.}
\add{We see utility in helping users make sense of multiple LLM responses. 
For instance, users may want to select the best option from among many, compose their own response through bricolage, consider many ideas during ideation, audit a model by looking at the variety of possible responses, or compare the functionality of different models or prompts.} 

\add{However, representing LLM responses at a scale necessary to see the distribution of possibilities also creates a condition where relevant variation may be hidden in plain sight: within a wall of similar text. One could turn to automatic analysis measures, but we constrain ourselves to showing the entirety of the text itself, as this does not constrict (by top-down design or button-up computation) which variations will be most useful to the user. }

Based on our \add{formative study} interviews with \add{a domain expert} end-user \add{(a doctor who uses LLMs to help them communicate effectively with patients)}, system designers \add{(a student building LLM-based tools and three startups founders/CTOs building LLM-backed products), model characterizers (doctors who are evaluating how well LLMs can answer medical questions)}, and model auditors \add{(researchers investigating how LLMs treat different sexual identity markers), we found that these} users typically engage in iterative inspection of 10s to 100s of LLM responses via a chat interface (which is slow) or by pasting responses into a spreadsheet (which is arduous and clunky). 
\cut{A number of emerging prompt engineering tools~\cite{promptfoo, weightsandbiases} 
support manual inspection as one piece of a larger evaluation environment, however they lack features for examining differences between 
groups of responses and focus more on automatic evaluation than sensemaking.}



This scale of 10s to 100s of LLM responses, where formative study participants reported spending time sensemaking and making decisions about model and prompt selection, is greater than the inspection of one to two outputs at a time (as is common in a chat or ‘playground’ type environments) but less than the many thousands of outputs typically involved in annotation studies. We therefore call this \textit{mesoscale}\
(“middle scale”) of LLM response sensemaking.


\add{In order to support users' sensemaking of LLM responses at the mesoscale, we designed and implemented several existing and novel text analysis algorithms and rendering techniques, each of which captures one or more aspects of LLM responses' possible variation and consistency. Established theories of human cognition describe how exposure to variation and consistency within prescribed structures can help humans more robustly form mental models of a phenomenon, e.g., how an LLM behaves.
Specifically, in line with Variation Theory~\cite{marton2014necessary}, the 
features we instantiate collectively identify patterns of consistency (Figure 1d, ``Exact Matches''—existing), variation (Figure 1c, ``Unique Words''—existing), or both (Figures 1a, 1b, ``Positional Diction Clustering (PDC)''—novel).
In line with Analogical Learning Theory~\cite{gentner1983structure}, PDC highlights positionally consistent analogical text across LLM responses such that users can see emergent 
relationships.  
\cut{The novel text analysis algorithm, Positional Diction Clustering (PDC), is designed to reveal positionally and semantically similar sentences within responses, taking advantage of the kinds of variation often seen in LLM responses.}
\cut{Our designs are informed also informed by literature on 
\cut{text visualization, }
how people skim text and make sense of variation across similar items, as well as prior systems supporting sensemaking of document corpora.}}

\add{By evaluating these interface features, we shed light on the answer to our research question:}

\begin{quote}
    \add{\textit{RQ: How can text-rendering interface design better support sensemaking of LLM outputs at the mesoscale?}}
\end{quote}

\noindent To understand \add{users' sensemaking support needs and how well these features did or did not serve them}\cut{potential utility of these features}, we ran a controlled user study and \add{a series of 8} open-ended case studies.
In the controlled user study our system was compared to a baseline interface where responses are presented linearly in groups based on the model or prompt; \cut{In these studies, we investigated two tasks: how the features support email rewriting with LLMs and how the features support comparing two different LLMs’ outputs.} \add{we investigated tasks that varied both in the number of outputs participants saw (from 9 to 50) and the kinds of sensemaking involved (an email rewriting task and identifying differences between models).}
In the case studies, participants were asked to use our system to support their own real world LLM uses. 
\cut{The case studies cover two quite different topics in LLM usage—identifying social bias and characterizing model capabilities for creative writing—and demonstrate how each feature has the potential to support open-ended exploration of LLMs.} 
\add{The case studies covered a range of tasks, including poetry and fiction writing, identifying social bias, and investigating LLM-generated legal advice, to name a few. }

\cut{We find that the features support detecting stylistic, content-related, and structural similarities and differences across responses, as well detecting diversity within a group of responses and outlier responses with unique features. Feature utility was a function of task type, the number of responses inspected, and user preferences for visualizations. }

\add{We report on a number of themes that emerged through these studies. These include the kinds of subtasks users perform (e.g. detecting stylistic versus content variation), approaches to these tasks (e.g. hypothesis confirmation), information processing styles (e.g. preferring an overview versus pagination through subsets), as well as user hesitancy to inspect too many responses at once (which was often overcome when exposed to our features).}


The contributions of this paper are:

\begin{itemize}
    \item \add{Formative studies that collect evidence of mesoscale text analysis of LLM responses for a variety of use cases.}
    \item \add{A controlled user study and open-ended case studies that demonstrate how our interface features can make sensemaking of LLM responses easier, and that many LLM-related tasks are intractable with current interfaces.}
    \item \add{A novel algorithm identifying similarities and variations across LLM responses, called Positional Diction Clustering (PDC), as well as a novel rendering for presenting the results of PDC.}
    \add{\item Design implications for future work on LLM response inspectors.}
\end{itemize}

\cut{In the discussion section, we consider how task-dependent variables impact feature utility, how feature implementation details may have limited their utility in certain scenarios, and how our evaluations, combined with relevant learning theory, motivate new directions for LLM interfaces.}
\add{In the discussion, we outline future design directions inspired by our findings, consider limitations to our approach to manual inspection, and reflect on the similarities and differences between designing for LLM response inspection versus other kinds of textual or machine learning data.}

\section{Related Work}
The objective of this work is to explore \add{both the tasks for which is it helpful to view many LLM responses at once in a single scrollable page, as well as the possible interface supports that could help users extract that utility.} 
\cut{how to support tasks that benefit from viewing
many LLM responses at once. } 
Here we review  precedents and influences on our work along three  axes: skimming support, text visualization, and sensemaking interfaces for generative AI.

\subsection{\add{Skimming and Skimming Support Tools}}

Reading and skimming are two distinct cognitive processes. Reading involves a sequential and comprehensive engagement with the text, whereas skimming is a strategic, selective, and non-sequential form of reading focusing on extracting the most salient information quickly~\cite{agosto2002bounded}. \cut{While reading ensures comprehension and retention of the content, studies have indicated a 
trade-off between reading speed and comprehension~\cite{masson1982cognitive, masson1983conceptual, tashman2011active, rayner2016so}. }
Skimming requires focused attention and strategic choices from the reader, which present additional cognitive challenges~\cite{duggan2009text}. 
Despite its challenges,
eye-tracking studies have found that skimming is very common, because of the time it saves~\cite{duggan2011skim, pernice2014people}.
In this work, we use a combination of word-, phrase-, and sentence-level highlights to \cut{visually promote}\add{call attention to}\cut{segments of} text that may be relevant to  sensemaking~\cite{pirolli2005sensemaking} tasks.

So far, the focus of skimming research has predominantly been on individual documents. However, our investigation diverges from this norm by emphasizing comparison across multiple documents. Traditional definitions might not classify such comparative reading as ``skimming,'' yet we posit that skimming operates as a fundamental cognitive process when readers contrast texts. In the subsequent section, we delve into studies centered around comparing documents. By consciously integrating skimming into the document comparison paradigm, we bridge these two realms of study, aiming to unveil innovative and more efficient interfaces for text comparison.


\subsection{\add{Cross-Document Comparison within Corpora}} \label{sec:corpora-related}
While skimming and reading strategies for a single document are well-established, generalizing to a collection of documents introduces the challenge of visualizing similarities and differences across documents~\cite{correll2011exploring}.
Many systems have been proposed towards this goal, but they either abstract away documents as a dot on a 2D plane (e.g., ~\cite{kim2016topiclens, le2019contravis}), as word pairs within a word cloud (e.g.,~\cite{diakopoulos2015compare, yatani2011review}), or as nodes within a graph (e.g.,~\cite{hu2016visualizing, hinrichs2015speculative}). Direct access to the text itself, \add{which is a precondition for the user to consider differences across texts, is often only possible as short document excerpts retrieved by a query containing a selected phrase or linguistic pattern (e.g., choosing a particular search term for text sliding~\cite{wordseer}),
hovering and clicking over abstract document representations to open a full-text view,
or drilling down into deeper layers of the interface.}
\add{In this work, we render all the documents (i.e., LLM responses) in their entirety as well as their relationships at a close textual level without requiring an initial query, as users may not yet know what an appropriate query would be, or even what they are looking for. This may be possible for LLM response inspection without overwhelming most users in large part because of the 
generative processes that produce the responses.}

One approach to visualizing text across documents involves leveraging the structure and layout of textual documents to aid skimming and comparison of text. 
\add{For instance, \textsc{VarifocalReader}~\cite{koch2014varifocalreader} supports skimming large complex text documents by using a multi-level layout where abstract summaries of varying detail of a document are shown alongside the document itself.}
The role of layout is important, especially in the early stages of interacting with a document, because readers often scan a document before delving into its details~\cite{leckner2012presentation,kress2020reading}; 
there is a human tendency
to treat words as “locations in space”~\cite{wolf2008proust}.\footnote{This is why people often remember where on the page a given piece of information was located, even if they cannot remember the information itself.}
Furthermore, the visual structure of documents can profoundly influence readers’ comprehension by affecting readers' assumptions, reading strategies, willingness to read, cognitive costs, and effort they must make to read~\cite{wright1999psychology, meyer1999importance, kendeou2007effects, olive2017processing}.   
\cut{In fact, when designed well, a document's layout can reduce cognitive overhead and guide the reader in understanding relationships between different document elements~\cite{britton1982effects}.} 
\add{We attempt to exploit the spatial aspects of document comprehension by decorating segments of text with highlights or changing their visual variables to show pre-computed cross-document relationships; this may recruit both visual and spatial memory and pattern recognition for textual sensemaking at scale.}

Text alignment is an often used strategy to facilitate direct comparison of the text across multiple documents, which can involve designing an algorithm for identifying shared patterns across texts and a method of visualising those shared patterns~\cite{yousef2020survey}. \add{For example, Tempura~\cite{wu2020tempura} uses linguistic features to find and summarize patterns in database queries, but it is limited to queries, which tend to be shorter and more structured than ordinary text.}  
\add{Prior work on text alignment has treated it as a sequence alignment problem and focused on algorithms involving edit distances and document-to-document matrices~\cite{dekker2011computer, meng2011determining}. CollateX~\cite{haentjens2015computer} later introduced variant graphs to enable the comparison and alignment of more than two documents and integrated the process into the digital collation workflow. However, the proposed alignment algorithms are still limited to single sentences with highly similar sentence structures~\cite{haentjens2015computer}. 
Gero et al.~\cite{gero2022sensemaking}'s preliminary work presented a sensemaking interface that uses concordance tables
to display LLM responses to support users in investigating problematic responses and distribution shifts across responses, but this preliminary work was not evaluated and was designed for single sentences rather than entire multi-sentence LLM responses.}

\add{Closely related work can also be found in the space of rendering code corpora, e.g., OverCode~\cite{overcode} and Examplore~\cite{examplore}, which pre-compute and render sub-document cross-document relationships using visual text attributes. Both OverCode and Examplore render entire corpora of text (code) with the same or similar purposes, i.e., a corpus of Python solutions to the same programming problem and a corpus of Java code examples mined from Github that all call the same API, respectively; our features render text generated by the same or similar types of models and prompts. OverCode and Examplore both abstract a minimal amount of the original code away; our features abstract away no text. OverCode precomputes line-level similarities and differences between corpus members and the most popular solution and reifies these differences using text saliency modulation; our features also precompute similarities and differences and reify them using visual text attributes. Examplore juxtaposes analogous components of code examples using a pre-defined template, visualized as a distribution over concrete readable code examples that form a coherent meta-example; our Positional Diction Clustering algorithm requires no pre-defined template to identify positionally and syntactically similar (analogous) sentences across many responses, and the interleaved rendering option also juxtaposes those analogous sentences like Examplore.}  



\add{LLM responses are a timely and important distinct type of corpus, and our features' designs have been inspired by---and, as in the case of PDC, necessarily unique from---prior systems' features that did not transfer from these other domains without significant insight.}

\subsection{Sensemaking Interfaces for \cut{Large Language Models}\add{Generative AI}}

\add{Generative AI models span multiple modalities, and interfaces for helping users understand and leverage their stochastic capabilities are in their infancy. Many systems in adjacent fields that aim to facilitate comparison among types of data other than text have adopted a grid view where information is organized along columns and rows. For instance, Mesh~\cite{chang2020mesh} helps consumers evaluate evidence about a product gathered across many different sources on a grid view, where columns are options to choose from and rows are criteria. Similarly, MLCube Explorer~\cite{kahng2016visual}, an interactive visualization tool for comparing machine learning results such as models’ performances over subsets of data, uses a grid view where each row represents a subset and the columns are summary statistics for each subset.  
In our work, the grid layout is specifically inspired from text-to-image generation “style guidelines” 
where AI-generated images are laid out in a grid that reflects controlled variations in the prompt~\cite{liu2022design}; in our work, the variation rendered in the grid is the result of multiple draws from the stochastic LLM, and, where indicated, different LLMs or variations on a prompt.}

The public availability, broad applicability, and performance of LLMs specifically has increased their adoption for a diverse range of applications that vary in domain and complexity. 
Prior work has shown that LLMs often generate long responses that negatively impact the user’s ability to understand and interact with the given output~\cite{jiang2023graphologue}. To mitigate this, there are interfaces to assist users with sensemaking and evaluating responses generated from LLMs. 
For instance, Graphalogue~\cite{jiang2023graphologue} and Sensescape~\cite{suh2023sensecape} transform LLMs outputs into diagrams that connect the concepts in an LLM response using a graph. These systems \add{enhance the rendering of individual LLM responses, rather than rendering the distribution over possible LLM responses to the same query, and are therefore do not explicitly address the stochastic nature of LLMs}. 
Meanwhile, Promptfoo~\cite{promptfoo} was developed for prompt engineering and evaluating prompts against predefined test cases; users can view prompts and inputs in a side-by-side display but the focus is on supporting automatic evaluation. In this work, we assume that users either do not have automatic quality measures for their task or that they do not yet know how to precisely define their goal. We seek to encourage inspection of outputs and do not abstract away from the text itself, as other prior very preliminary work has done~\cite{gero2022sensemaking}.

\section{Formative Interviews}
\cut{In order to confirm that manual inspection of outputs was a key part of working with and developing systems with LLMs,} 
\noindent We interviewed eight people working with and developing systems with LLMs\cut{working with LLMs about their process}. In semi-structured interviews, we asked interviewees about their process for working with or developing LLM-based systems, how they selected pre-trained and/or fine-tuned models, and how any prompt engineering was structured. The interview guide can be found in Appendix \autoref{app:formativeinterviewguideline}. 

\subsection{Participants}
\add{Participants were recruited through the authors' professional networks.}
Our interviewees included two doctors investigating LLMs in medical contexts, one researcher investigating the creative abilities of LLMs, three start-up founders or CTOs (from three different companies) who oversee the development of LLM-based public-facing products that support writing, and two artist-researchers who build and interrogate chatbots’ attitudes towards queer identities.


\subsection{Findings}
Overall, we found that everyone we interviewed engaged in manual inspection of outputs. This happened at different scales: from comparing two or three outputs \cut{to looking at around 25,}to reading a list of 1000 outputs. \add{(As described in the introduction, we refer to the middle of this range as the \textit{mesoscale} for text analysis.)} Sometimes it involved clearly-defined annotation, but often it involved discussions among the system designers or with users. All interviewees had developed ad-hoc processes to support this inspection. Many noted that they put outputs in a spreadsheet, as this facilitated both sharing of outputs as well as increasing their readability. 
In addition, three main themes emerged:

\subsubsection{Failure of Automatic Evaluation}
All interviewees said that automatic evaluations were not \cut{very} useful when it came to developing LLM applications. Several interviewees had explicitly investigated if benchmark evaluations of \cut{pre-trained models} \add{LLMs} correlated with which model worked best for their use case and found no correlation. Because automatic evaluations could not predict success at their task, all interviewees engaged in some kind of manual inspection for evaluation.

\subsubsection{Sensemaking of LLM Responses}
Several interviewees discussed a sensemaking process with LLM responses. The artist-researchers discussed reading and comparing outputs as a key part of their development process, to understand if a model they were training was “getting better” or to compare how models treated heterosexual vs.~homosexual couples. The researcher working on creative writing applications discussed both the prompt engineering he did to develop the system as well as the prompt engineering he saw his users engage in, e.g., users comparing model capabilities for prompts in first v.~third person. One of the start-up co-founders mentioned a variety of kinds of manual inspection that involved sensemaking, from panel discussions with users to internal evaluations where “we rely on our own literary skills to evaluate the outputs.” \cut{He said that typically they looked at 1-25 outputs at once.} Another start-up founder described their process as putting a few people in a conference room with a lot of outputs and having them read them all. 
In particular, he noted the importance of detecting problematic outlier outputs, because preventing these was key to developing and maintaining user trust.
These observations show the importance of sensemaking to the use of LLMs and the range of forms sensemaking takes.

\subsubsection{Complex or Nuanced Annotation}
The third start-up founder described their process as printing out a thousand outputs and reading them all manually, noting which ones “worked” for their intended use case and which did not, and using this ad-hoc, manual annotation to select an appropriate fine-tuned model. Both 
doctors we interviewed were engaged in research projects evaluating LLMs in a clinical setting. One doctor described the detailed and highly-skilled human annotation involved in evaluating LLMs, and noted that the same grading metrics used for evaluating doctors were being used to evaluate LLM responses. 
In each of these cases, the analysis is complex, beyond the standard affordance in chat UIs, i.e., "like" / "dislike" buttons.


\add{\section{Feature Design: Existing and Novel Sensemaking Support Features}}
We designed and prototyped \add{instantiations of several existing and novel algorithms and renderings for scaling up LLM response sensemaking.} \add{Each highlights or juxtaposes words, phrases, or entire sentences based on their relationship to the entire collection of LLM responses.}\cut{via manual inspection of LLM responses} 
\add{Each interface feature is a combination of a text analysis algorithm and a rendering technique}. \add{We view these features as \textit{tech probes}:} \cut{an initial exploration of}\add{a non-exhaustive set of points in} a design space for LLM response inspectors \add{that scale up human inspection}. 

\subsection{Design \add{Goals}}

\subsubsection{Scale up the number of LLM outputs a human can consume.}

\cut{We wanted to support the inspection of responses from an LLM at the \textit{mesoscale}:} We aim to make 10s to 100s of LLM responses cognitively comfortable to peruse, \add{as this was the scale we found to be most relavent in our formative study. This is the same scale at which Examplore~\cite{examplore} was evaluated.} This range is relevant for a variety of tasks. When writing an email, a writer \add{may have an easier time recognizing tone and diction variation
across 10 different LLM responses.} When looking for inspiration, a designer may look at 50 responses to surface diverse possibilities. When 
checking for outlier responses, a system developer may want to look at 100s of responses. 

\subsubsection{\cut{Engaging with Text Itself, Rather than Abstractions}\add{Show the entire collection of LLM outputs as text---not abstractions}}

\cut{Additionally, we want to support engagement with the text itself.} 
\add{In our formative study, we found that automated analysis rarely captured what the participants were looking for when inspecting LLM responses.} \cut{Automated analysis that abstracts away from the text would foreground some aspects of text while failing to encode others that may be important to a user.
Additionally, nuance can be lost, and 
users may or may not be able to notice and recover from algorithmic or AI ``choices'' they disagree with.}
\cut{We want users to feel confident that they are being shown}\add{The choice to have minimal textual abstractions in prior publications like OverCode~\cite{overcode} served users well: users could recognize what \textit{they} thought were good and bad aspects of programming composition in student solutions which were worthy of comment. Similarly, we think users in our context should be able to see all of the responses 
and perform their own sensemaking. } 

\subsubsection{\add{Do not require users to select text ``lenses'' with which to see the data}}
\add{Some text analysis tools require users to select (or accept recommended) search terms in order to access the most powerful text analysis algorithms and renderings~\cite{wordseer}. However, in our formative study, participants seemed to prefer engaging with the text directly without having to articulate a lens with which to look at the corpus, since their analysis goal may be initially under-defined. For this reason, we want to instantiate flexible features that allow users to immediately inspect the entirety of LLM responses without requiring the user to choose or accept a particular lens (e.g., search term) with which to render them.} 

\subsubsection{\add{Show pre-computed relationships within the rendering of the text itself}}
\add{Like prior work on rendering sub-document cross-document relationships within a textual corpus~\cite{overcode, examplore}, we want to decorate text to show pre-computed relationships, such as string matches across responses. In this way,}
we help users shift cognitive bandwidth away from identifying \cut{similarities}\add{overlapping or ``unique'' language} to answering more complicated questions.
Additionally, we create skimmable visual patterns across all of the responses.


\subsubsection{Support \add{a variety of} sensemaking \add{sub-tasks for both system designers and system end-users}}
We want to support a wide range of tasks that involve sensemaking. For example, we want to support the detection of similarities and differences between individual responses as well as groups of responses, and support the detection of  “outlier” responses (or parts of responses). In alignment with sensemaking literature \cite{pirolli2005sensemaking}, the goals of user tasks could be very open-ended 
or be hypothesis driven
depending on the task itself and how “far along” a user is in the sensemaking process.

\subsection{Relevant Theory} 
By interacting with a system, users may develop a mental model that guides their future interactions with it~\cite{staggers1993mental}. Variation Theory~\cite{marton2014necessary} and Analogical Learning Theory~\cite{gentner1983structure,gentner2013analogical} each propose mechanisms for how people may conceive and update their mental models based on concrete examples, or use their mental model in new situations. 
Extensive evidence congruent with each theory has been collected in many domains~\cite{vtInTheWild15}, though not yet for the task of mentally modeling LLMs \add{or mentally modeling the space of possibilities, e.g., in an ideation task, generated by an LLM.} \add{One prior piece of HCI work, ParaLib~\cite{paralib}, does explicitly exploit these theories for system feature design, but does this in the domain of code: example-based programming library comparison and selection.}

\add{Variation Theory describes how helping people perceive the different dimensions of consistency and variation across examples (here, LLM responses) of the object of learning helps them more quickly and robustly leap to more accurate mental models. Analogical Learning Theory describes how people can form mental models or schema from perceiving structural analogical relationships across superficially varying examples (again, here LLM responses).  In this work, in line with Variation Theory, the existing and novel features instantiated and described in the next subsection collectively identify patterns of consistency,
variation,
or both;
they are explicitly designed to make emergent dimensions of consistency and variation easier for the user to perceive. }

\subsection{Feature Descriptions}
\label{sec:feature-descriptions}

We \cut{build our system}\add{implemented our LLM response sensemaking features within a fork of the} open source project, ChainForge~\cite{arawjo2023}, which is a visual programming language for generating, inspecting, and analyzing LLM responses.\footnote{\add{ChainForge, and the text rendering functionality we built atop it, is written in Typescript. Our code imports the ChainForge-retreived LLM responses as a list of JSON objects and renders them.}} \cut{It contains functionality to query one or more LLMs at once with the same prompt as well as easily generate prompt variations, e.g., using “Who invented the \{object\}?” and specifying a list of objects to be substituted in to generate prompt permutations; the output of those LLM queries can then be piped into visualizations. Our features are implemented as such visualizations.}
\add{The features are a combination of text analysis algorithms and rendering. The initial two algorithms and first layout are conceptually straight-forward extensions of existing features, while the final novel algorithm and its corresponding custom layout are designed specifically for analyzing and rendering collections of LLM outputs that are, by construction (e.g., multiple draws from the same model), variations on a theme.}


\subsubsection{Exact Matches}

This conceptually straight-forward feature enables users to see how similar responses are by finding and identifying “exact matches” across responses.  There are many ways to implement this functionality. We detect and highlight the longest common substrings, as that appeared to be most robust to a wide variety of response types during informal evaluations. 
\cut{\paragraph{Connection to Theory}}
The hypothesized benefit of this feature is that users can shift cognitive bandwidth from identifying \cut{similarities}\add{overlapping language} to answering more complicated questions.
Additionally, it gives a skimmable sense of patterns of repetition across all of the responses.

\begin{figure}[h]
\centering
\includegraphics[width=.9\textwidth]{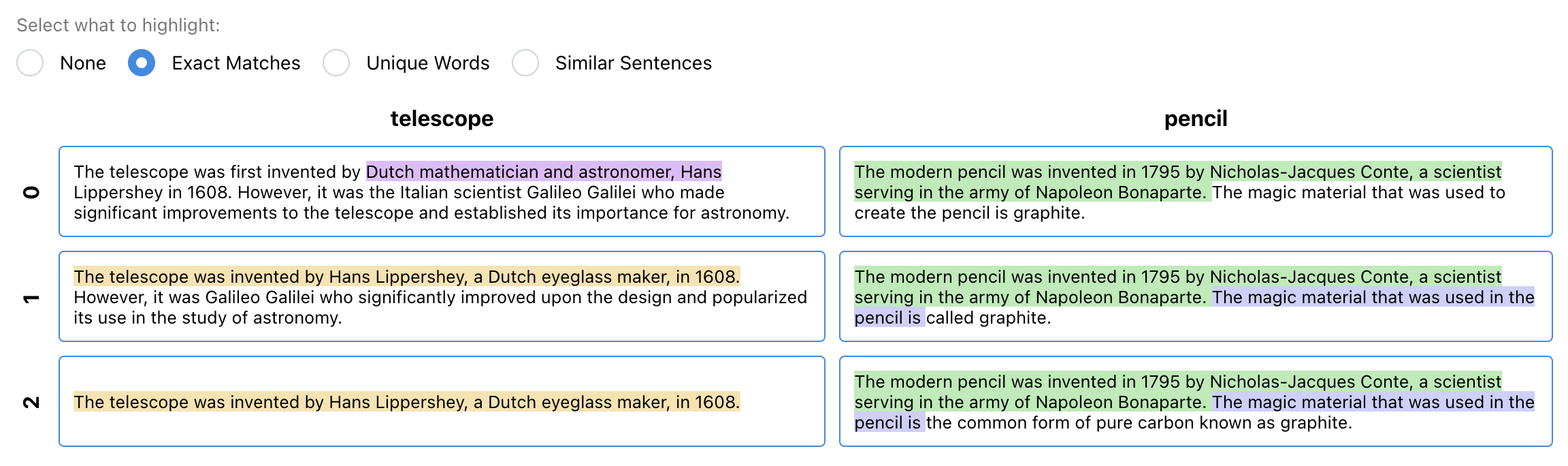}
\caption{Example of the `exact matches' feature for the prompt ``Who invented the \{object\}?" where the objects are `pencil` and `telescope' and each prompt had $n = 3$ generations. Exact matches makes it easy to identify portions of responses that are matching across multiple responses.}
\Description{Screenshot of the exact matches interface. There is a grid of six responses. The columns are labeled `telescope' and `pencil'. The rows are labeled 0, 1, 2. There are four colors for the highlighting, with each color highlighting one to 3 sentences that are exact matches.}
\label{fig:features-em}
\end{figure}

\paragraph{Algorithm and 
Rendering}
We identify substrings to highlight in five steps:
(1) find common substrings in pairs of responses, (2) split substrings that span sentence boundaries, (3) filter short substrings, (4) rank them based on a combination of substring length and how many responses they appear in, and (5) retain the top $k = \text{min}(12, |\text{responses}|/2)$. For details see Appendix \autoref{app:exactmatches}.
Each set of exact matches is highlighted in its own color; examples are shown in the Grid Layout (\autoref{sec:gridlayout}) in \autoref{fig:teaser}(d) and \autoref{fig:features-em}.

\subsubsection{Unique Words}
This feature allows users to see what is distinctive about responses by highlighting “unique” words in each response. There are many ways to measure uniqueness; we use a simple measure that has been widely studied in Natural Language Processing and Information Retrieval: term frequency-inverse document frequency (TF-IDF)~\cite{jurafsky2023speech}. TF-IDF uses the frequency of a word in a given document as well as its frequency across all documents to calculate a score for how ``representative'' a word is of a document relative to the rest of the collection.

\cut{\paragraph{Connection to Theory} 
Like exact matches, the hypothesized benefit of this feature is that
we are freeing up cognitive bandwidth for more complicated questions. And, like exact matches, pre-computed unique words supports the skimming of differences between responses, letting users skip over similarities. 
}

\begin{figure}[h]
\centering
\includegraphics[width=.9\textwidth]{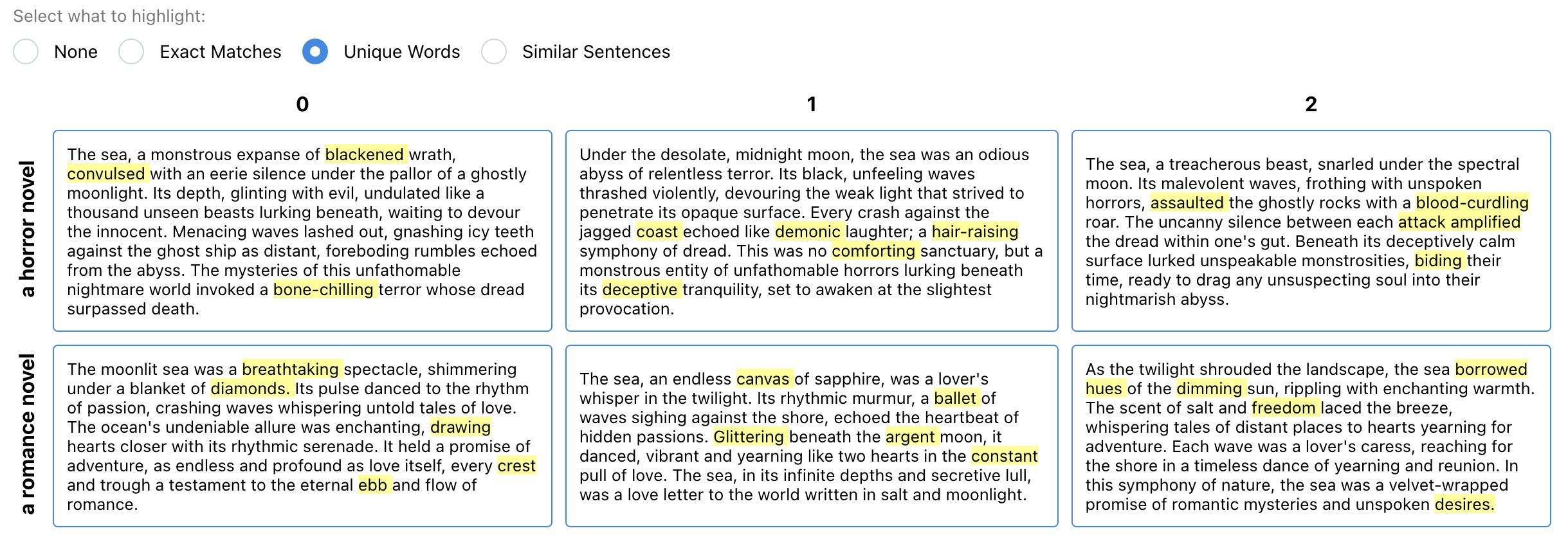}
\caption{Example of the `unique words' feature for the prompt ``Write a short paragraph about the sea in the style of \{style\}." where the styles are `a horror novel` and `a romance novel' and each prompt had $n = 3$ generations. Unique words makes it easy to see how word choice is influenced by the style.}
\Description{Screenshot of the exact matches feature. There are six cells. The columns are labeled 0, 1, 2. The rows are labeled `a horror novel' and `a romance novel'. Two to five words are highlighted in each cell, all in the same color.}
\label{fig:features-uw}
\end{figure}

\paragraph{Algorithm and Rendering}
We calculate TF-IDF values using the Wink-NLP Javascript library~\cite{winknlp}. The set of documents is defined as the set of responses generated by the selected LLM(s). For instance, if the system was run with two models, three prompt variations, and three responses for each combination, there are 2*3*3=18 responses, which comprise the set. Term frequency is calculated for each response. We remove stop words\cut{, as we found that they were less meaningful for many tasks}.
We highlight the five top-scoring unique words in each response. See \autoref{fig:teaser}(c) and \autoref{fig:features-uw} for examples, shown in the Grid Layout.

\subsubsection{Positional Diction Clustering (PDC)}

\cut{When looking at many LLM responses to the same or similar prompts, one often begins to notice an emergent template in terms of the sequence of the content. For instance, responses may start and end with very similar sentences. Sometimes content is slightly reordered, or the LLM interjects an extra sentence or leaves one out.
But there remains an emergent structure across responses.}
\cut{Our exact matches feature does not reveal these patterns because there are variations in how the emergent templates are expressed.}

\cut{The general idea for this feature is to}This novel algorithm is designed to identify, when present, any emergent structure \add{(and variation within that structure)}---by finding groups of sentences from different responses that are similar in textual content \textit{and their location within responses}.
Each group will then correspond to one part of an emergent template. \add{Orphan one-off sentences in single response which do not correspond to the content and position of sentences in other responses are preserved as singletons.}

Theories of human concept learning suggest that a key step in \add{forming accurate, robust mental models of a phenomenon} is to be able to discern the underlying dimensions of variation (Variation Theory) and any latent structures beneath superficial details (Analogical Learning Theory). 
By detecting and communicating which sentences are both structurally analogous and semantically similar, users should be able to more easily identify emergent structures, as well as compare and contrast particular structural elements that may vary in meaningful ways across responses. These theories assert that these subtasks are key ingredients in forming those robust accurate mental models, i.e., learning from the LLM outputs in order to better perform their overarching task. 

\begin{figure}[h]
\centering
\includegraphics[width=.9\textwidth]{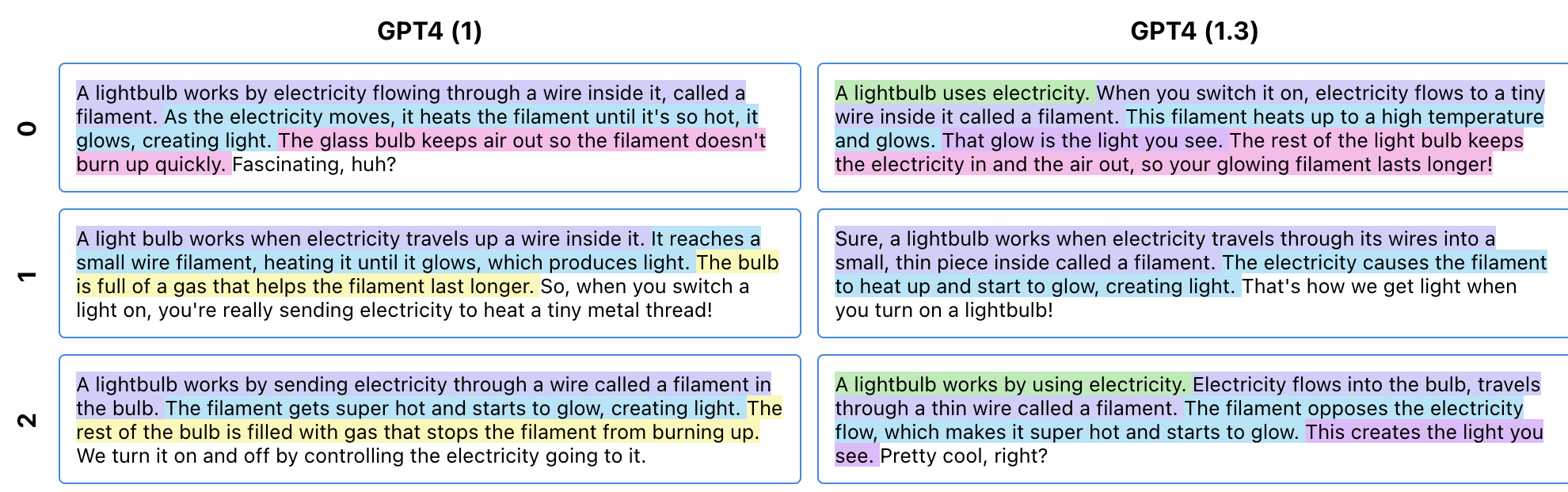}
\caption{Example of the PDC feature in the grid layout for the prompt ``Explain how a lightbulb works to a 12 year old." for GPT4 temperature=1 and GPT4 temperature=1.3. In the grid view, structurally and semantically similar sentences are highlighted in the same color; notice that the sentences highlighted in yellow are both about how gas supports filament longevity.}
\Description{}
\label{fig:features-ss}
\end{figure}

\begin{figure}[h]
\centering
\includegraphics[width=.9\textwidth]{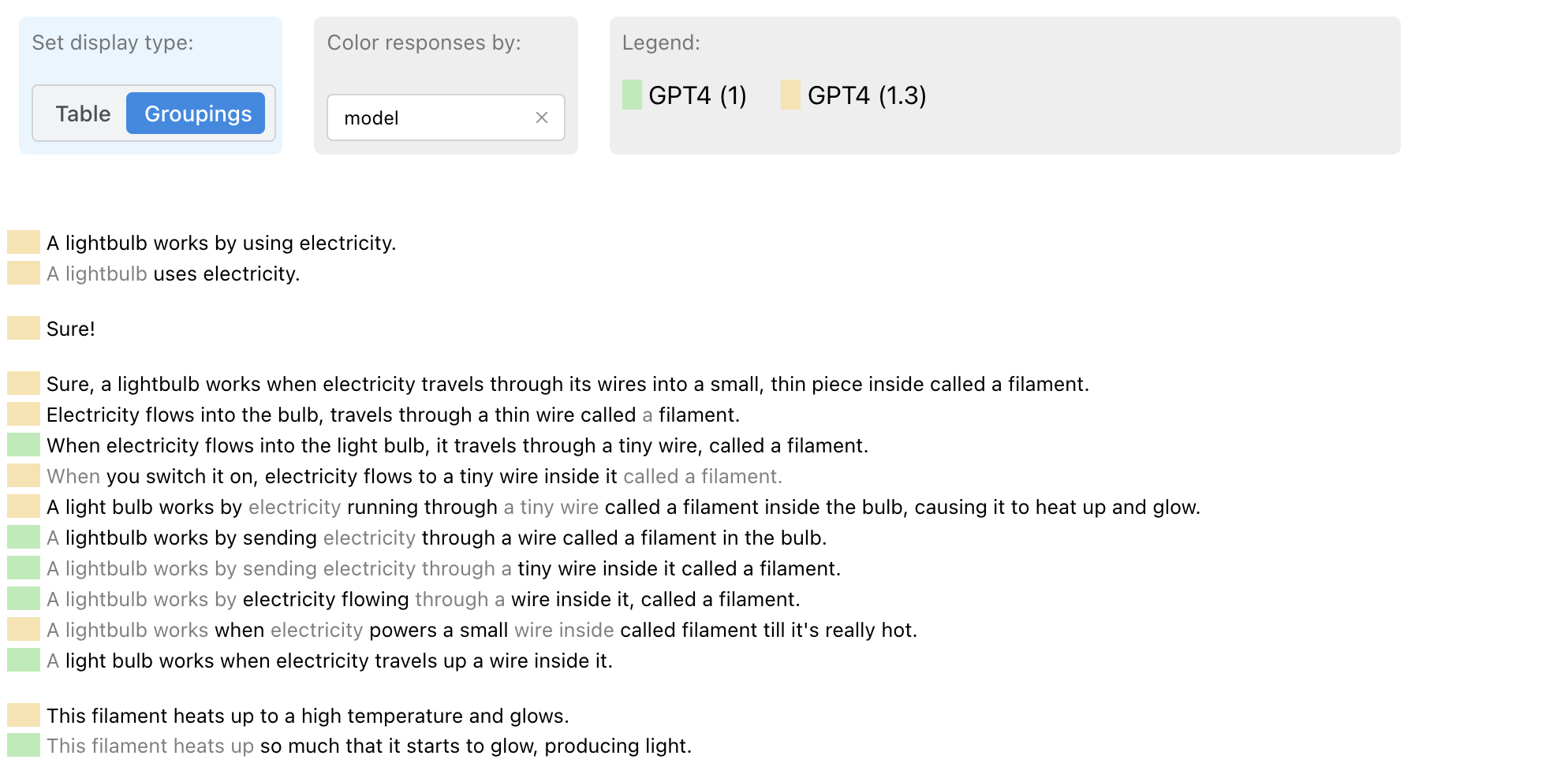}
\caption{Example of the PDC feature in the interleaved layout for the same prompt, model, and temperature settings as in \autoref{fig:features-ss}, i.e., ``Explain how a lightbulb works to a 12 year old." for GPT4 temperature=1 and GPT4 temperature=1.3. In the interleaved view, structurally and semantically similar sentences are grouped, with the color \add{patch to the left} indicating which model version produced them; notice that all the opening `topic' sentences are \add{shown together with redundant text grayed out.}}
\Description{}
\label{fig:features-interleaved}
\end{figure}

\paragraph{Algorithm and Rendering} \label{alg-grouping}
We create groups of sentences using a form of single-link agglomerative clustering.
For every pair of sentences, we calculate their content similarity as exact diction overlap normalized by their combined length.\footnote{This is a slight variant of Bray–Curtis Similarity. For details see the Appendix \autoref{app:pdc}}
Initially, every sentence is placed in its own group.
Then, we iterate through all sentence pairs, in decreasing order of content similarity.
For each pair, we merge the groups that contain those sentences if (1) the two sentences are sufficiently similar in content and normalized location in the LLM responses\footnote{See the Appendix \autoref{app:pdc} for the calculation, threshold values, etc., used in user study} and (2) merging creates a group where at least 70\% of sentences are all from different responses, \add{since we are interested in analogous \textit{cross}-document sentences}.
Finally, for each group, we calculate the mean normalized location in the LLM response of sentences in the group. \add{Groups of one are permitted; these capture sentences that have no analogous sentence (in terms of both content and approximate normalized location) in any other response.}
The algorithm returns (a) all the groups\add{, including groups of one,} and (b) the mean position for each group. \add{This algorithms' results can be rendered in one of two ways: using the same color to highlight all the sentences that share a group, as is done in the Grid Layout (shown both in \autoref{fig:teaser}(a) and \autoref{fig:features-ss}), and listing sentences, by group, in order of their groups' normalized mean location within the responses, as is done in the Interleaved Layout (\autoref{sec:interleaved}), which is specifically designed as an alternative layout for the results of PDC (shown in \autoref{fig:features-interleaved}).} 

\cut{\subsubsub{Visualization 1: Grid PDC}
Within the Grid Layout, we highlight sentences from the same group in the same color (\autoref{fig:teaser} and \autoref{fig:features-pdc}).
This allows users to visually distinguish the components of the emergent template.}

\subsubsection{Grid Layout}
\label{sec:gridlayout}
LLM responses are laid out in a grid, with user-defined variables for the columns and rows. \add{For instance, users can select that the different models queried determine the columns, and repeated generations from the same prompt determine the rows.} This view allows users to see many responses with controlled variations (model, prompt, temperature, etc.) side by side. For example, in \autoref{fig:teaser}, the template prompt asks the model to generate a short story. \cut{for a five year old child about a creature that loses something and then finds it again.} Here, prompt variations---different possible story characters---define the rows, and the $n=3$ different responses per prompt define the columns. 
There may be more than two variables a user is interested in, for instance also comparing models. In the top right of the interface, the user must select which value of the remaining variables to surface.
\cut{In \autoref{fig:teaser} the user has chosen to see responses from GPT-4.}
Currently, the grid does not support viewing more than two variables (i.e. the column and row variables) at a time, though extensions could allow this. 

\cut{\paragraph{Connection to Theory}}
There are two hypothesized benefits of this view. 
One is based on an understanding of human perception: 
the grid layout should help users compare more LLM responses because the spatial arrangement assists their memory. 
The other benefit is based on Variation Theory, which posits that discerning the impact of a critical aspect, for example model temperature, is only possible when experiencing variation along that dimension, isolated from variation along other dimensions. 
The user-configurable grid layout makes it possible for users to isolate dimensions of variation, enabling them to discern \add{their qualities and critical values, for instance how model version affects the language of its responses to the same prompt.}

\subsubsection{Interleaved Layout (enabled by PDC)}
\label{sec:interleaved}
\cut{All of the previous features are modifications of the text in the grid layout.
The interleaved layout is an alternative to the grid, in which} Sentences from different responses are strategically interleaved.
The rendering (\autoref{fig:teaser}(b) and \autoref{fig:features-interleaved}) is generated by printing out the groups produced by the algorithm.
Groups are ordered based on the average position calculated by the algorithm.
This means they roughly follow the flow of the original prompt.
Each group is rendered with one sentence per line, in an order that maximizes exact word overlap between adjacent sentences in the group.
If a sentence has any exact word overlap with the sentence above, that is if the $i^{th}$ word in both sentences is the same, the word in the sentence below is grayed out.
A small amount of whitespace separates each group.
A colored square to the left of each sentence  indicates which prompting condition generated it.

We hypothesise that this rendering will (1) be particularly useful for users who wish to remix parts of several different LLM responses, (2) support the identification of rare response components, as they will be easily identified as singleton groups, and (3) support characterizing the complete distribution of LLM responses, since the volume of screen real estate taken up by any given group is proportional to the number of responses that include a sentence in that group.

\section{User Study}

\cut{\subsection{Methodology}}

We instantiate these features \cut{---grid layout, exact matches, unique words, grid PDC, and interleaved PDC---}into a single interface, which we call the “exploratory interface”. \cut{This represents an initial probe that can inform future designs.}
{We also implement a baseline interface that represents a status quo in LLM output rendering: \add{a linear list of outputs, as one might get from an API call or pasting outputs into a spreadsheet, as some formative interviewees did}. The baseline interface has two additional capabilities: listing outputs grouped by model or prompt, and putting each group in a collapsible container, such that users can have as many groups open at a time as they like. Within each group, responses are presented linearly top-to-bottom. \cut{These two features are collectively referred to as "linear".} See Appendix \autoref{app:baselineinterface} for a screenshot.

\add{We ran a controlled user study that investigated when and how the different features were helpful relative to the baseline interface. This allowed us to investigate: \textit{Which features best support sensemaking tasks over many LLM responses?}}
\cut{, we ran two controlled user studies to investigate how well each feature supports two tasks: a task where participants are asked to select the best response from a set of 9 responses, and a task where participants are asked to determine differences between models by looking at a set of 50 responses (25 responses per model for two models). }
\cut{We designed this study to investigate the follow questions:}
\cut{
\begin{enumerate}
    \item{\add{Which features best support sensemaking tasks over many LLM outputs?}}
    \item{\add{How does the collection of features compare to a baseline interface for such tasks?}}
\end{enumerate}}
\add{The user study protocol put participants in two different scenarios (writing an email, comparing two models in order to choose one) at two different scales (10 and 50 LLM responses, respectively). The exact prompts for each scenario can be found in the Appendix \autoref{app:prompts}.\footnote{Although narrowing our participants to students is not necessary to answer our research questions, by narrowing the pool to students we were able to pick tasks that were more likely to be personally compelling and realistic.}  
This study evaluates these features' utility for end-users who attempt these tasks.}
\footnote{\add{In the case studies we investigate how more targeted user groups (like researchers and system designers) use the features in self-selected and self-directed tasks.}}



\subsection{Participants}
Participants were recruited from a local university via mailing lists and flyers. Participants had to be over 18 years old, fluent in English, and a student of some kind (undergraduate, masters, or PhD). 
We recruited 24 participants (11 women, 12 men, and 1 non-binary). Eight participants were 18-24 years old, and the remaining 16 were 25-34. All were graduate students of some kind (15 Ph.D. students, 9 Master’s students).
In the pre-study survey, we asked participants about their confidence in writing important email and their experience with LLMs. Details can be found in Appendix \autoref{app:participantdata}.  Most participants felt very confident writing important email and had some experience with LLMs. Few participants had explored the differences between LLMs before.

\subsection{Study Procedure}
All studies took place in person. Participants completed the tasks using the facilitator’s laptop. Facilitators were authors of the paper. The general flow of the study procedure is rendered visually in \autoref{fig:studyflow}, and is \cut{presented briefly here in text} \add{described in more detail in Appendix \autoref{app:userstudy1procedure}.} The exact prompts can be found in the Appendix \autoref{app:prompts}.

\begin{figure}[h]
\centering
\includegraphics[width=.8\textwidth]{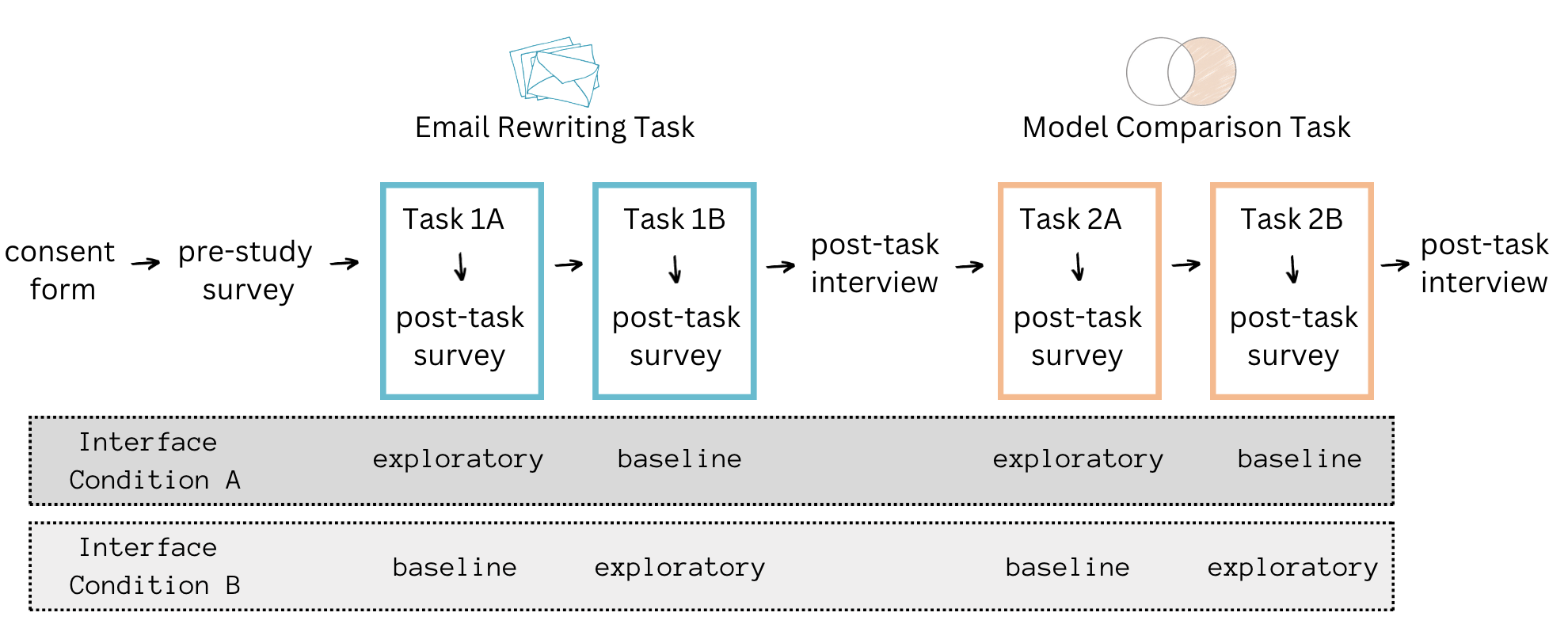}
\caption{Study Process: each participant performs two email rewriting tasks with different UIs and two model comparison tasks with different UIs. Interface conditions are counterbalanced.}
\label{fig:studyflow}
\Description{Flow chart of how the study proceeds. From left to right: consent form, right arrow, pre-study survey, right arrow, task 1A and then post-task survey, right arrow, task 1B and then post-task survey, right arrow, post-task interview, right arrow, task 2A and then post-task survey, right arrow, task 2B and then post-task survey, right arrow, post-task interview. Below this flow are two boxes one on top of the other. The first is labeled Interface Condition A. The second is labeled Interface Condition B. In condition A participants get the exploratory interface in tasks 1A and 2A and then the baseline interface in tasks 1B and 2B. The reverse is true for condition B.}
\end{figure}

\cut{Participants were walked through informed consent, and then audio and screen recording began. Participants then filled out a short demographic survey, including questions about their exposure to LLMs. Participants then went through two different tasks---the email rewriting task and the model comparison task---doing each task twice, once in each condition. For the first task, before each interface condition, participants were shown a short tutorial introducing the interface's features.

Within a given task, e.g., email rewriting, the order of the scenarios remained the same and \textit{the interface conditions were counterbalanced}. After completing the task in each scenario, participants filled out a short survey, which included close-ended questions about cognitive load, how realistic the task was, and how well they believed they had performed on the task. After each task, the facilitator conducted a short semi-structured interview with open-ended questions about the utility of the features in both interfaces.
At the end of the study, the facilitator stopped the recording, allowed participants to ask questions about the study, and conferred study payment.
See Appendix \autoref{app:surveys} for survey and interview questions. }

\subsection{Task 1: Email (Re)writing}
We chose an email (re)writing task as prior literature has used this task with LLMs~\cite{buschek2021impact, goodman2022lampost} and we found this is a common use case of LLMs~\cite{reddit1, reddit2}. 
\add{Additionally, we hypothesized that users may benefit from seeing multiple variations at once, rather than one at a time (as would be the case in a chat interface) as users may better be able to compare and contrast responses to select the best one, and may be more able to recombine elements of different versions into their final draft.}

Participants were shown nine different LLM responses that each rewrote the same initial email draft. \cut{rewriting an initial rough draft of an email for the same relatively high-stakes scenario.} The prompts used in the study were chosen to be realistic \cut{and high-stakes} for the recruited participants (university students): asking a professor for a recommendation (Task1A) and asking an internship manager for a later start date (Task1B). 
\cut{A user, if they were consulting an LLM for this purpose in real life, would probably consider their word choices carefully.}
Participants were asked to first select the LLM response that was closest to the one that they would send in real life. 
We gave participants up to 3 minutes to complete this part of the task, given that pilot participants typically did not take more than 2.5 minutes.
Then, participants were asked to edit their chosen LLM response to make it closer to what they would want to send. We mentioned that participants could also remix parts from different LLM outputs. 
We gave participants 2 minutes for this part of the task, as most participants could complete it within this time and the outlier participants would be prevented from editing for too long.

\subsection{Task 2: Model Comparison}
In this task, participants were asked to compare models; they looked at 25 responses from GPT-3.5 and 25 from GPT-4. \cut{The same prompt was given to both two models.} \add{We selected this task to be one that users may want to engage in (compare model behaviour) and hypothesized that it may be quite difficult with status quo tools.}
There were two \cut{scenarios} \add{prompts, one for Task2A and one for Task2B}: asking for advice about how to skim a book and asking for advice on how to prepare in the week before an important final exam. 
The participants' task was to list as many differences as they could between the two models' responses. 
This task is quite open-ended. In pilot studies, participants needed at least 5 minutes to come up with even a few model differences. However, some participants found the task exhausting and would ask to give up around 8 minutes. For this reason, we gave participants 10 minutes for this task, and \cut{invited} \add{allowed} them to stop early if they felt they had completed the task to the best of their ability.

\subsection{Analysis}

\subsubsection{Quantitative Analysis}
\cut{We used the study video recordings to manually determine the amount of time each participant required to finish each task, and then} We ran statistical tests to compare responses to all the Likert scale survey questions as well as time on task. 
\cut{Specifically, we used the Mann-Whitney U test for nonparametric data to compare the results of Likert scale questions, and a two-tailed t-test for data with similar variance to analyze time on task data. }
\cut{For both tasks, we first looked for learning or scenario variation effects. For example, for the email rewriting task, was the first email rewriting scenario more difficult than the second? 
If no learning effects were found, interface conditions were compared within subjects across both scenarios. If learning effects \textit{were} found, interface conditions were compared between subjects for each scenario separately.
We looked for significant differences in: self-reported cognitive load, self-reported task success, and interface preferences, as well as, in the model comparison task, their actual success, i.e., number of differences found.\footnote{In the email rewriting task, self-reported success is essentially the same as actual success as there is no objective metric for how well their email matches what they would really send.} }
In the model comparison task, to determine how many differences were found by each participant, we first put all participants' listed differences into one list and had one author, who was blind to participant and condition, clean the data manually.\footnote{This was because some participants would write paragraphs instead of bullet points, or include two differences in a single bullet point.} Then we use counts of how many differences each participant wrote down in each condition.

\subsubsection{Qualitative Analysis}
Two authors followed a general inductive approach for analyzing qualitative data~\cite{thomas2006general}. Both listened to all interviews to gain context, and, via transcriptions, pulled quotes relevant to the research questions. With a shared set of quotes, the researchers independently came up with codes for the quotes, then came together to discuss and create a codebook. This codebook was shared with other members of the research team, discussed and revised. With the revised codebook, the two authors re-coded half of the data, and disagreements were discussed and codes were revised based on discussion. Finally, the remaining data was re-coded by a single author.



\begin{table*}
  \caption{Codebook.}
  \label{tab:codes}
  \small
  \begin{tabular}{lp{12cm}}
    \toprule
    Code & Definition\\
    \midrule
    \multicolumn{2}{l}{\textbf{Kinds of Similarities and Differences Noticed}} \\

        style
        & Stylistic or tone similarities and differences.
        \\
        content
        & Content similarities and differences, including counting phrases and the length of responses.
        \\
        granularity
        & Level of abstraction of similarities and differences, e.g. high level v. low level or  “granularity”.
        \\
        structure
        & Structural similarities and differences between responses, including ideas of responses being ‘segmented’.
        \\
        outliers
        & Identifying outliers or unique features.
        \\
        diversity
        & Determining consistency or diversity across sets of responses. 
        \\

    \midrule
        
    \multicolumn{2}{l}{\textbf{Preferences about Interface Elements}} \\

        display
        & Preferences about viewing all or many responses at once.
        \\
        scroll
        & Preferences about needing to scroll to return to responses.
        \\
        options
        & Preferences about the ability to disable features.
        \\
        delineation
        & Preferences about delineation between prompt groups (e.g. responses to different prompt variations).
        \\
        visual learner
        & Preferences about learning or processing information visually.
        \\

    \midrule
        
    \multicolumn{2}{l}{\textbf{Elements of Reading Process}} \\

        speed
        & What made reading responses faster or slower.
        \\
        read all
        & Reading the entirety of groups of responses (as opposed to skimming).
        \\
        skim
        & Skimming groups of responses. 
        \\
    
    \midrule
        
    \multicolumn{2}{l}{\textbf{Cognitive Elements}} \\
        
        memory
        & Issues of working memory, forgetting response content or location, or needing to reread to recall responses.
        \\
        ease
        & Cognitive demand of different interface elements.
        \\
        overwhelm
        & Feelings of overwhelm or stress, especially in response to inspecting too much information at once.
        \\
        exhaustion
        & Feelings of exhaustion or “zoning out”.
        \\
        focus
        & Ability (or inability) to focus on specific responses or groups of responses.
        \\
        difficulty
        & Feelings of task difficulty, including not knowing how to start a task.
        \\
    
    \midrule
        
    \multicolumn{2}{l}{\textbf{Methods of Detecting Similarities and Differences}} \\

        confirmation
        & Performing “hypothesis testing” or otherwise confirming or verifying an idea.
        \\
        comparison
        & Directly comparing responses by having them nearby or side by side.
        \\
        absent color
        & Making use of unhighlighted text segments.
        \\
        
    \midrule
        
    \multicolumn{2}{l}{\textbf{Feature Accuracy and Understanding}} \\

        accuracy
        & Determining how accurate or trustworthy different features are. 
        \\
        not understanding
        & Not understanding how a feature worked or why it performed a certain way.
        \\

  \bottomrule
\end{tabular}

\end{table*}

\add{\autoref{tab:codes} shows the results of our analysis of the interviews with participants. These themes reflect a wide range of participant responses, from subtasks performed (e.g., detecting diversity of responses) to methods for performing these subtasks (e.g., confirming hypotheses) to the cognitive elements involved in performing them (e.g., focus and working memory). We use these themes when analyzing how participants did or did not make use of the features in each of the tasks in the user study, as seen in the subsequent two subsections, but also present them as a result of the study that can inform future system designers and user studies, as they reflect a range of methods, issues and concerns that come up when users inspect LLM responses.}

\subsection{Email (Re)writing Results}


\subsubsection{Few Quantitative Differences Between the Exploratory and Baseline Interface}
\add{Participants completed task 1B faster than task 1A, and felt less rushed. For this reason, we split out analysis comparing interface conditions between tasks 1A and 1B; further details can be found in Appendix \autoref{app:emailrewritingresultsdetails}.}
Participants in both interface conditions and scenarios were quite successful at the task, with all participants rating their final edited email as above a 5 on a 7-point scale where 7 was “I would definitely send this email.” There were no significant differences between the interface conditions for either task 1A or 1B for how successful they were at the task (how likely they were to send the email) nor how long they took. This suggests that the exploratory features did not obviously impact participants’ ability to do the task, perhaps because the task was easy enough to achieve near perfection in both interfaces.
When asked which of the two interfaces was easier and which was more overwhelming, there was no clear preference; participants varied in which interface they preferred. 

\add{\subsubsection{Qualitative Differences Between the Exploratory and Baseline Interface} 
Despite in aggregate there being few differences between the interfaces, there did exist preferences that varied across participants.}

\add{In the exploratory interface, both the layout and multiple different highlighting features were explicitly called out as helpful, relative to the baseline interface. Some participants appreciated being able to see all responses at once:  \textit{“I think that's what's nice about the grid; [the emails] are all right there. And so your eyes can sort of flutter and dart around as you're reading each one”} (P9). Some participants appreciated how the highlighting features allowed them to easily compare segments of responses (using PDC in the Grid Layout) and identify stylistic choices that matched their own style (using the Unique Words feature). These participants also noted that, in the baseline interface, needing to scroll to revisit responses was challenging because \textit{“if you want to reference the first email that you read versus the ninth, you have to scroll up and then you lose the view of the ninth one"} (P9).}

\add{In contrast, other participants said that the exploratory interface felt overwhelming because seeing all nine at a time \textit{“was too much information. … The [baseline linear view] was more organized”} (P1). In particular, the baseline interface allowed users to collapse groups of responses\cut{(which were grouped by prompt)}, allowing participants to focus on one group of responses at a time. Participant responses to the two interface conditions seemed to be influenced by their information processing style, something we saw come up in the case studies as well, and we discuss in further detail in the Discussion section.}



\subsubsection{Utility of Different Features}

\cut{In this section we go over the features individually, reporting on how participants used (or did not use) each feature, including the baseline interface feature, which we call the “linear” feature. After the thematic analysis of the interviews, we had a codebook which represented what participants were concerned about and their actions taken with the different features. In the Appendix, Table X shows our codes, which are used for both the Email Rewriting task as well as the Model Comparison task.}
\add{Most participants spent little time with any of the features, although some participants spent significant time in Grid PDC and Unique Words; details can be found in Appendix \autoref{app:emailrewritingresultsdetails}. The four participants who spent the majority of their time in the Grid PDC feature noted that because the email messages had a common structure, the highlighted sentences allowed for easy visual segmentation of the responses 
--- this allowed participants to easily compare across the individual segments. One participant said that the Interleaved PDC feature matched the way he currently used LLMs for writing tasks, except he normally had to do it by manually copy and pasting sentences from multiple responses grouped by their semantic function, such that he could pick the best sentence per group. }
\add{Two participants spent the majority of their time in Unique Words feature. One of these, P7, said that this feature made it \textit{“easier for me to just look and see what makes each response unique.”} }
\add{Overall, participants found the Grid PDC most useful, but it also appears that this task may have been too easy to require much detailed exploration of the responses, as participants could read the entirely of all nine in just a few minutes. The next task proved more difficult, and therefore highlighted the utility of the features when users are more likely to struggle with existing interfaces.}

\subsection{Results for Model Comparison Task}


\subsubsection{Quantitative Differences Between Interface Conditions}
\cut{First we checked the survey questions and time on task between the first and second model comparison scenario and found no differences.} 
When comparing the interface conditions, we found no difference between the cognitive load questions. However, we did find that participants reported being more successful in finding differences with the exploratory interface than with the baseline interface (p < .05, Mann Whitney U task for nonparametric data).
When participants were asked to rate which interface made the task easier and which was more overwhelming, participants reported a strong preference for the exploratory interface, as can be seen in \autoref{fig:modeldiff-interface}.

\begin{figure}
     \centering
     \begin{subfigure}[b]{0.4\textwidth}
         \centering
         \includegraphics[width=\textwidth]{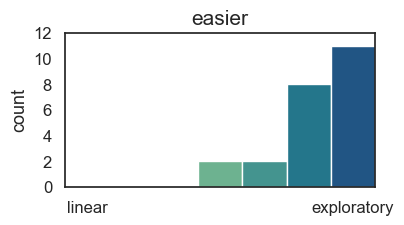}
         \caption{``Which interface made the task easier?''}
         \label{fig:modeldiff-success-baseline}
     \end{subfigure}
     \hfill
     \begin{subfigure}[b]{0.4\textwidth}
         \centering
         \includegraphics[width=\textwidth]{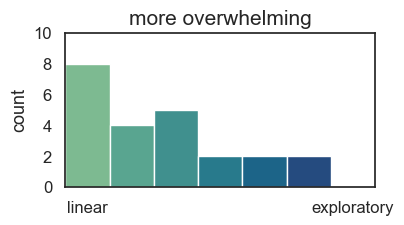}
         \caption{``Which interface was more overwhelming?''}
         \label{fig:modeldiff-success-exploratory}
     \end{subfigure}
        \caption{Participant interface preferences for the model comparison task.}
        \label{fig:modeldiff-interface}
        \Description{Left: histogram titled `easier' with 7 buckets from `linear' to `exploratory'. Histogram is heavily skewed towards `exploratory' on the right; from left to right the counts are 0, 0, 0, 2, 3, 8, 11. Right: histogram titled `more overwhelming' with 7 buckets from `linear' to `exploratory'. Histogram is heavily skewed towards `linear' on the left; from left to right the counts are 8, 4, 5, 2, 2, 2, 0.}
\end{figure}

Participants reported feeling more successful with the exploratory interface, and reported that the exploratory interface made the task easier, but did they actually come up with more differences? 
\cut{First we checked if participants wrote down more differences for the second scenario. 
We found no significant differences in the number of differences found in the first and second scenario. Then, we checked if the interface condition impacted the number of differences found. Looking at both tasks together (first and second) we find no significant difference. However, if we compare conditions between tasks, we find a significant difference for the first time with the task but not the second time, see \autoref{tab:modeldiff}.}
\cut{The significance testing above indicates that participants got the most benefit from the exploratory interface if they saw it first.}
\add{Since we may reasonably expect that participants may get better at this task the second time around, we ran an} ANOVA analysis of how task order and interface condition impact the number of differences found. Our dependent variable is the number of differences found, our fixed effects are task order and interface condition, and our random variable is participant ID. We find that the task order is not significant but the interface condition is (p < 0.01), demonstrating that the exploratory features did help participants find more differences.


We found that participants spent significantly less time with the baseline interface than the exploratory one (baseline: 8.8 minutes; exploratory: 9.7 minutes; p < .05; two-tailed t-test for data with similar variance). Since participants also found more differences with the exploratory interface, it’s reasonable to assume that there may be a correlation between time on task and number of differences found. We calculate the Pearson correlation coefficient and p-value for testing non-correlation, one-tailed where we test for positive correlation, and find that time on task and number of differences is significantly positively correlated (p < 0.05). \add{This indicates that the exploratory interface may have allowed participants to perform the task better through increasing engagement, allowing participants to stay on task longer.} 

\add{\subsubsection{Qualitative Differences Between Interface Conditions} In the baseline interface, because participants had to scroll up and down to get to responses from the two models, many participants reported forgetting what they had determined about a single model. 
P2 suggested this was \textit{“because we have a limited memory window,”} such that skimming through the linear view resulted in forgetting what they were even trying to discern. P13 described the exploratory interface as: \textit{“We can literally see the difference, what this group says versus what that group says. So it gives a side by side comparison."}}

\add{All participants reported that the exploratory interface made the task easier because it allowed for easier recognition of similarities and differences between the two sets of responses. Participants used the features in the exploratory interface in a variety of ways, some of which we had imagined and some of which we did not; details can be found in the next section. While this may be an expected result, we did not see this result in the email rewriting task, suggesting that the utility of LLM inspector interfaces may be dependent on the number of responses users need to inspect (in these studies, 50 LLM responses versus 9, the former of which is more firmly at the mesoscale text analysis) or something about the task itself. In the next section we dig into the variable utility of different features to support this task.}

\subsubsection{Utility of Different Features}

\add{\autoref{fig:modeldiff} shows the time participants spent with each feature, as well as how many participants indicated a given feature was useful. Grid PDC was the most used and most highly preferred feature. Participants found that Grid PDC help them read responses faster: \textit{“If I saw a similar sentence highlighted then I didn't read the sentence completely, I knew that I already heard them before”} (P1). Additionally, this feature helped people identify output segements; P5 said that \textit{"in the [baseline], I had to actually segment everything in my head after reading it. And in the [exploratory], I didn't have to spend much time reading ... because the segmentation was already being performed."} While we expected participants would use Grid PDC to notice similarities based on what was highlighted, we didn't expect participants to also consider what wasn't highlighted, as P7 did, who said \textit{"if it's not color coded at all then it's probably a unique thing."}}

\add{Interleaved PDC was the second most preferred feature, along with Exact Matches. Interleaved PDC was mainly helpful in detecting content differences between the two models. P1 described that 
\textit{“there were certain clusters which were only present in one of the models.”} Similarly, participants noted that singleton groups indicated that only one model came up with that suggestion.
P15 also used Interleaved PDC to detect structural differences between the two models: \textit{“It would help me to find out that there's some major difference in the distribution, like GPT3.5 had occupied a bunch of them at the top. 
... 
I found out that it actually shows more introductory sentences.”} }

\add{All participants used Exact Matches in very similar ways: to determine how consistent or diverse a model was. Still, this was very useful, and participants spent a decent amount of time in this feature. Unique Words was the least preferred feature, though participants did spend time with it. No participants mentioned in the interview a preference for Unique Words or how they used it.}

\cut{Here we review how people thought about and used the various features, including the baseline interface feature which we call the “linear” feature. This section reports on the results of the thematic analysis (codebook can be found in the Appendix in \autoref{tab:codes}) as well as analysis of how much time participants spent with the different features (see \autoref{fig:modeldiff-featuretiming}) to identify themes in how people used the different features.}


\begin{figure}
     \centering
     \begin{subfigure}[b]{0.4\textwidth}
         \centering
         \includegraphics[width=\textwidth]{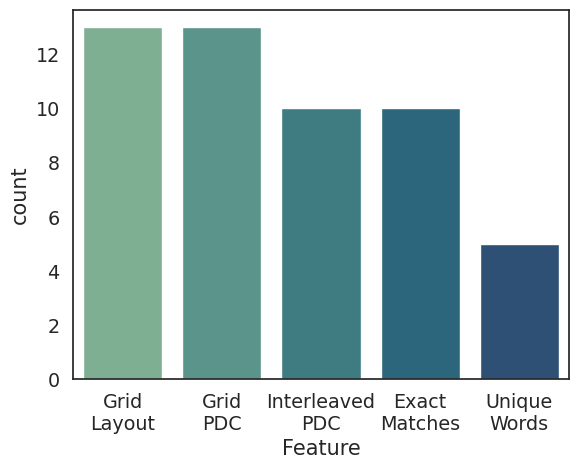}
         \caption{\add{How many participants marked a given feature as useful in the post-task survey. Here `Grid Layout' refers to the general laying out of responses in a user-defined grid.}}
         \label{fig:modeldiff-1}
     \end{subfigure}
     \hfill
     \begin{subfigure}[b]{0.4\textwidth}
         \centering
         \includegraphics[width=\textwidth]{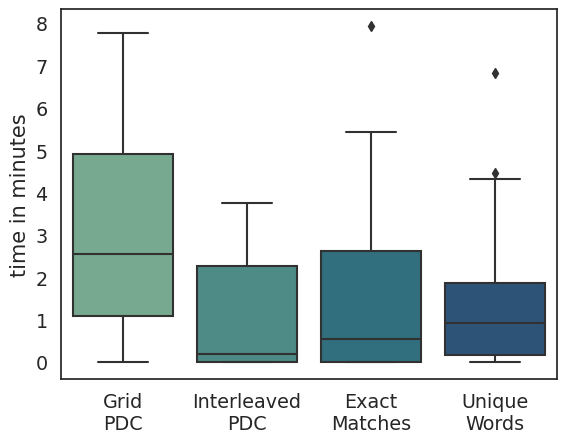}
         \caption{\add{Time spent in each feature across all participants, ordered by how many participants marked a given feature as useful in the post-task survey, i.e., counts from Figure 8a to the left.}}
         \label{fig:modeldiff-2}
     \end{subfigure}
        \caption{\add{On the left we show the `popularity' of features based on how many participants marked that feature as useful in the post-task survey. On the right we show how much time participants spent in each feature. Although participants reported equal preference for Interleaved PDC and Exact Matches, on average they spent more time in Exact Matches.}}
        \label{fig:modeldiff}
        \Description{}
\end{figure}


\section{Case Studies}

The controlled user study investigated how participants used the features in two different tasks, including variation across participants. However, participants had no control over the prompts, models, or what they were attempting to do. This section reports on \add{eight} case studies, where we recruited participants via our professional networks who were interested in understanding and examining LLM outputs in \add{the context of} their own work. Participants were asked to bring their own tasks, and attempt those tasks using the exploratory interface. \add{\autoref{tab:casestudies} shows for each case study the domain of the task, a short description of the task(s) performed, as well as the max number of responses viewed at any one time and the approximate number of words per response. We attempted to recruit a wide range of case studies, in terms of the kinds of tasks but also the anticipated number and length of responses inspected.}

\begin{table*}
	\caption{\add{Details of the participants in the case studies.} }
  \label{tab:casestudies}
  \small
  {\renewcommand{\arraystretch}{1.6}
  \begin{tabular}{lp{2cm}lcc}
    \toprule
    ID & Domain & Task/s  & \thead{Max Num. \\ Responses} & \thead{Approx.\\  Words per Response}\\
    \midrule

        P1 & creative writing & \makecell{generating insightful connections; \\ generating character voices} & 20 & 500
        \\

        P2a, P2b & \makecell{model auditing; \\ intersectional AI} & \makecell{how models treat identity markers; \\ how models understand historical events} & 39 & 100 \\

        P3 & creative writing & \makecell{story continuation; \\ poetry writing} & 40 & 50 \\

        P4 & journalism & \makecell{prompt engineering for journalistic feedback} & 50 & 250 \\

        P5 & academic writing & \makecell{prompt engineering for writing task; \\ improve section outline} & 20 & 150 \\

        P6 & business & business idea generation & 50 & 350 \\

        P7 & law & identifying trademark confusion & 20 & 200 \\

        P8 & history & \makecell{how agency is represented in responses to \\questions about historical events} & 100 & 100 \\

  \bottomrule
\end{tabular}
}

\end{table*}

The case studies were open-ended, with a facilitator (one of the authors) giving participants a tutorial of the system \add{for 10-15 minutes} and then the participants interacting with the system in whatever way they were most interested in. Participants were asked to share their screen and to think aloud as they worked. The facilitator asked questions during the study to illuminate the participants’ goals, findings, and struggles\add{; the interview guideline is in Appendix \autoref{app:casestudyinterviewguideline}.} 
Case studies lasted between one and two hours. 

\add{We report two kinds of results. The first is a set of themes that emerged about the utility of the different features. These themes were determined by two authors watching the case studies, writing descriptive summaries of each, and then collecting and collating themes iteratively. The second is an analysis of how user preferences, task, and the number and length of responses informed differences across the case studies. This analysis was done via repeated conversations with all paper authors, all of whom read the descriptive summaries of the case studies and two of whom had watched the video recordings. During our analysis, we recognized that similar themes identified for the controlled user study (see \autoref{tab:codes}) were emerging. This demonstrated the generalizability of these themes.}



\add{\subsection{Utility of Different Features}}


In this subsection we go over the features and how they were useful (or not) in different contexts. Features could perform quite differently in different contexts, where the combination of a participant's goal and the way the feature performed on the responses they were inspecting accounted for variation in utility. 

\add{\subsubsection{Exact Matches} As in the controlled user study, participants used Exact Matches mostly to identify how consistent or diverse a set of responses was (i.e. more exact matches indicated more consistent). However, participants were sometimes unsure if there should be many exact matches or not. For, instance P1 asked the facilitator if there should be so few exact matches. Across most of the tasks there were few exact matches, and the matches tended to be syntactic phrases (such as ``The relationship between'' or ``However, it is important to note'') rather than semantic ones. \cut{P3 did not expect there to be many, if any, exact matches because their responses were short.}}

\add{One exception to the trend of few exact matches came in P8's use case, where they looked at the Falcon7B ~\cite{falcon40b} model's response to a question about the industrial revolution. In this case, there were many sentence-length exact matches. This meant they were able to use the un-highlighted text to identifier outliers in responses.
Most participants worked with an OpenAI GPT model, so it may be that some models have less variation---and more exact matches---than others. With P8, for the same prompt there were far less exact matches when prompting GPT4 than Falcon7B.}

\add{\subsubsection{Unique Words} Several participants used Unique Words as a kind of extractive summary of a response. P1's first task was asking a model to identify how two unrelated concepts might be related, with the intention of getting the model to generate ``insightful connections.'' As soon as he turned on Unique Words he commented that the highlighted words might be considered ``intersecting'' words between the two concepts,
and subsequently used this feature to skim through responses to quickly identify what connection(s) the model had generated.
P6, who had a model generate business ideas in response to a prompt about how to support the circular economy, noted that Unique Words allowed her to quickly 
identify what the generated business idea was about or its general domain (e.g., fashion, or food packaging).}

\add{Other participants thought that Unique Words could summarize a response, but found that, as implemented, it did not quite do what they wanted. P7 prompted a model to identify potential ways two trademarks might get confused.
She thought that Unique Words could be useful to skim the ``argument'' of the response, but determined that it didn't really do this. To consider a very different context, P3, after having a model generate a poem in the style of a particular poet, wanted Unique Words to highlight ``strong choices'' or ``interesting phrases'' rather than just words that were unique, as unique words were sometimes circumstantially unique rather than unique in a way that related to the task at hand.}

\add{\subsubsection{Grid PDC} As in the controlled user studies, Grid PDC allowed participants to determine consistency or uniqueness
across groups of responses in a way that went beyond exact matches. P1, when having a model take on different character ``voices'' (such as answering a question as if it were the philosopher Hegel or a secret service MI6 Officer) used Grid PDC to immediately and visually tell whether or not the model had a strong sense of what that character was like, because that set of responses would have similar colors highlighted. If the model was generating a distinct character voice, P1 would then use the highlights to determine what characteristics were present by looking at what the clustered sentences were about. In a very different context, P5 was trying to determine if his chosen model was capable of doing the rewriting task he had prompted,
where the goal was to have the model rewrite a paper section outline to be more clear and concise. He noted that, outside the case study, when he uses an LLM with a chat interface, he is often trying to determine if the model is incapable of doing the task or if the problem is one of prompt engineering. Grid PDC helped him view at a glance, by looking at around five responses to each of several prompt variations, whether or not the model was capable of the task at hand. P7, looking at ways two trademarks might get confused, used Grid PDC not to skim over semantic variation but syntactic variation. She 
would focus on a particular cluster and look at the variation across responses, some of which would use different adjectives which could make a big difference when adjudicating trademark confusion.}

\add{However, Grid PDC did fail if there was too much variation. P3, looking at story continuation and poetry generation, had nothing highlighted in this view because there was so much variation across the generated responses. Another failure mode was the opposite: so much was highlighted that it was difficult to identify clusters.}

\add{\subsubsection{Interleaved PDC} Similar to the controlled user study, Interleaved PDC was useful to identify smaller variations within a cluster of similar sentences. P5, who had prompted the model to rewrite an outline to be more clear and concise, talked about this view as ``shopping cart'' mode, where he could pick the best version of a rewritten sentence. P2b, looking at how a model understood certain historical events and historical writers, particularly enjoyed how this view allowed her to see repeated phrases and syntactic patterns that occurred across responses which, had the model known more about the events or people, should have been more distinct. P6, looking at business idea generation, thought this view was useful for noticing themes, as each cluster showed responses that had a similar business idea.}

\add{The main issue with Interleaved PDC was a lack of context. P7, looking at trademark confusion, 
felt she needed to see sentences in context in order to understand what they really meant.
In contrast, P5, asking the model to rewrite an outline, felt context was not quite as necessary, perhaps because the task was list-like and therefore each sentence was not as dependent on the surrounding sentences. P2b, looking at historical events and writers, similarly was unworried by lack of context, perhaps because she was less interested in the utility of a particular response and more interested in reading  ``across'' responses in order to understand generally how a model would respond to certain prompts.}

\add{\subsection{Differences Across User Attitudes, Tasks, and Number and Length of Responses }}

\add{\subsubsection{User Hesitancy with Many Responses} Several participants were hesitant to look at too many responses at once, generally because they assumed it would be too overwhelming. This was the case with P4, who had been doing prompt engineering via an API; P5, who was familiar with a chat interface and regenerating only a few times to test prompts; and P8, who looked at four responses per prompt and would use automatic analysis 
for larger numbers of responses.} 

\add{All of these participants, when the facilitator suggested looking at more responses, immediately saw the value and began to attempt new kinds of tasks. For instance, P5 noted that he could put together a better single response by inspecting and selecting portions of many responses. He said, \textit{``the more the better, because I have more ways to choose from...the more I can explore.''} P4 noticed that her prompt failed 1 out of 50 times, which was important as she wanted to incorporate the prompt into a web application where incorrect outputs would break downstream functionality. P6 realized that she could inspect up to 50 business ideas at a time, and then drill down into subsets of ideas around a common theme to identify the most innovative ideas, which would be impossible to do with fewer responses.}

\add{Not all participants had this initial hesitancy. Some participants had an existing practice inspecting many responses, such as P3 who often looked at 100s of responses when using an LLM to generate creative writing. Others, such as P1, P2a, and P2b, had wanted to look at many responses at once but had never had an appropriate interface to do so. These participants were interested in model capabilities generally, rather than using a model to do a particular task. It may be that users interested in \textit{understanding} models, rather than using them, may more easily understand the utility of our system, whereas users with specific tasks may need demonstration to see the potential for many responses to be useful.}

\add{One participant retained their hesitancy even after experiencing our exploratory interface. P7, looking at trademark confusion, could not imagine looking at more than 20 responses at a time, and thought 10 was an appropriate number. P7 also had the least experience with LLMs of all our participants.}


\add{\subsubsection{Task Stage and User Preferences Impacted How Many Responses to View at Once} Even participants who wanted to inspect 100s of outputs would sometimes want to only see a few at a time. A quintessential example was P3, working on creative writing, who wanted to go over 100s of outputs but felt seeing them all at once was overwhelming.
This is a reflection of the information processing style preferences we saw in the email rewriting task,
where some participants wanted to see all the emails at once and others wanted to be able to view just three at a time.}
\add{While information processing style may be driving user preferences here, it was also the case that some participants wanted to view fewer responses after first having inspected many more. For instance, both P1 and P6 noted that viewing many responses allowed them to notice which subsets of responses were most relevant to their task, and then dig into this subset.} 

\add{\subsubsection{Response Attributes Required Algorithms be More Adaptive} Some participants looked at very short responses, just one or two sentences; P3 even suggested wanting to look at sentence fragments. Others looked at responses with five or more paragraphs, getting up to 500 words per response.
P4 wanted to look at outputs that were returned as a JSON array, a very distinct syntactic format we had not considered. 
As these attributes changed, the algorithms needed to respond to them appropriately. For instance, the PDC algorithm segmented responses based on a sentence parser, which was not appropriate for responses only one or two sentences in length, nor JSON outputs. As responses got longer and the number of responses got higher, the number of clusters detected by PDC increased, often resulting in too many clusters to highlight in distinct colors. With shorter and smaller numbers of responses, our Unique Words algorithm would often highlight less meaningful words, as the TF-IDF metric had insuffieicnt data for patterns to emerge.}

\section{Discussion}

\add{Overall, we saw that the existing and novel features we designed can support LLM response sensemaking. Our novel algorithm, PDC, and both the Grid and Interleaved renderings, were particularly helpful for a variety of tasks, and often the most popular feature with participants, indicating the value of algorithms and renderings that are designed specifically for LLM responses. As LLMs are increasingly adopted, 
supporting end-users, system developers, and system examiners in making sense of the stochastic capabilities of LLMs becomes an increasingly important area of study.} 
\add{Given that the features implemented in this work 
are in line with design implications of Variation Theory and Analogical Learning Theory, the results suggest that there may be further utility of these theories for guiding the design of future systems that help users make sense of data and form mental models from examples.}

\cut{Most existing work on ways to analyze text data is based on tasks such as reviewing different translations of the same source document or comparing separately written documents about the same subject (or programs all solving the same problem). 
LLM responses are a timely and important distinct type of corpus, and our features' designs have been inspired by---and, as in the case of PDC, necessarily unique from---prior systems' features (reviewed in \autoref{sec:corpora-related}) that did not transfer from these other domains without significant insight.} 

\add{\cut{We implemented our features based on Variation Theory and Analogical Learning Theory; these theories have been applied to the analysis of code and text/data, but not to textual output generated by LLMs. Code and non-LLM generated text tend to not have as many variables as the LLM context, where prompt input, model choice, and temperature, for example, influence and introduce variation. Therefore, the analysis of these domains do not need to be as modular or deal with as much (and different kinds of) variation. Work in the machine learning context is, in some ways, more similar to the LLM context, in that there are variables (e.g. kinds of inputs) and ways to slice the data (e.g. different biases to look at) that introduce variation. However, these outputs, often numerical data or much shorter, simpler text data (like user queries as in Tempura~\cite{wu2020tempura}), have different properties than that of LLM outputs, which can be longer and reflect more nuanced variations that existing approaches do not address.}} 




\add{
\subsection{Design Implications and Future Work}}

\add{We report on design implications that arose from our studies. The suggested features were either explicitly mentioned by participants or developed through observation of how participants used and responded to our exploratory interface.} 

\add{\subsubsection{New Algorithms and Renderings} When participants discussed the utility of the features, they often suggested new algorithmic goals or additional renderings that would better support their tasks. For instance, while participants consistently used Exact Matches to determine how consistent or diverse responses were, participants also commented that the highlighted phrases were mostly stylistic or syntactic phrases, rather than semantic or content-heavy ones. In this way, they wanted a different feature, one that did not focus solely on exact matches but rather one that took into account other features in the text. 
Below we list a variety of new algorithm goals for highlighting words or phrases:}

\begin{itemize}
\item \add{Select words or phrases unique to a general linguistic corpus (rather than to the set of LLM responses).}
\item \add{Select phrase-length ``fuzzy'' matches, rather than exact matches.}
\item \add{Select only exact matches which contain ``content'' words, as opposed to phrases that are more stylistic.}
\item \add{Allow users to select a single PDC group and remove the highlighting of other groups to improve visual focus.}
\item \add{Show descriptive characteristics of the PDC algorithm output, such as number of sentences per group.}
\item \add{In the Interleaved PDC view, allow users to see the response from which a given sentence came from.} 

\end{itemize}

\add{We also saw a need for applying algorithms over subsets of responses. For instance, instead of highlighting words that are unique to a single response, highlight words that are unique to the set of responses from a single model or prompt. All our algorithms (and future algorithms) could be applied to, for instance, rows or columns, which would allow the algorithms to reflect meaningful variation across user-selected subsets of responses.}

\add{\subsubsection{Support User-Defined Queries} Although we did not want users to have to pre-determine a specific ``lens'' (i.e., search term) through which to view the data initially, after interacting with our features many participants wanted to customize the algorithms in some way. 
For instance, participants wanted to define a phrase on which to search for ``fuzzy'' matches, or define the part-of-speech an algorithm focused on. Another route would be to let users write their own algorithm, which could then be applied to the responses. In a way, our features represented good ``defaults'' that let users determine what kind of queries they would like to create or customize.}



\add{\subsubsection{Support Response ``Subsetting''} Many users wanted to ``drill down'' into a subset of the responses, or select or save responses such that they could view just these ``good'' ones without having to sort through the rest. Interface features to support this kind of dynamic inspection of responses would allow users to move along the sensemaking process, narrowing the responses they are investigating. A very simple version of this is allowing users to collapse or hide columns or rows of responses; more sophisticated versions include letting users mark some responses to ``store'' and letting them hide all unmarked responses at a later point, or letting them dynamically hide or show individual cells.}

\add{\subsubsection{Support Explicit Annotation} \label{sec:annotation} Future work could use our exploratory interface as the basis for an annotation tool, where annotations could then be exported for analysis or used to subset the data. This would allow users to partake in more structured tasks while retaining the utility of the highlighting features. This could also make response inspectors useful in the 1000+ response scale, where users typically focus on annotation rather than sensemaking, but still could use the support of algorithms and renderings.}

\add{\subsubsection{Integrate Automated Analysis into Response Inspectors} Some participants noted that they do use automated analysis as part of their sensemaking process. For instance, they may apply sentiment analysis to LLM responses and then inspect responses according to which sentiment bucket they fall in. We see potential for integrating such analysis into response inspectors where, for instance, responses could be colored according to how they are classified by an algorithm, or reordered in the grid according to their classification. 
We still firmly believe that, in sensemaking processes, users must be able to look at the raw text itself as automated analysis may hide or fail to capture variation that users would be interested in if they knew it was there, but visually incorporating automated analysis could be the best of both worlds: users can make use of automated analyses to direct their attention, but retain closeness to the text.}

\add{\subsection{Limitations}}

\add{\subsubsection{Limitations based on algorithm and rendering implementations.} As we consider our work to be a tech probe to better understand how to support the inspection of many LLM responses at once, the algorithms and renderings could be improved. 
There were small problems with our implementation. Occasionally Exact Matches would highlight, e.g., a three-word and four-word phrase in different colors, although the three-word phrase was a subset of the four-word phrase. Sometimes the PDC groups got too large, resulting in it not being clear why two sentences were clustered together. On the rendering front, users noted that some colors in the highlighting were too similar to distinguish. Although these problems decreased the utility of the features, we did not observe a huge impact on our results.}

\add{\subsubsection{Limitations to our studies, and future directions for study.} The controlled user study explored only two tasks. While our case studies attempted to test a wider range of tasks, they could not dive deep into any one task. Future work could more precisely investigate the utility of LLM response inspectors for specific tasks, such as auditing models for harmful content, end-user selection of a preferred LLM, or prompt engineering for system designers. Another angle would be to investigate particular domains, such as the use of LLMs in legal or medical contexts.} 

\add{A final angle for future studies would be testing at larger scales, e.g., 1000 responses at once, which is 10 times the upper limit of what case study participants examined for tasks they brought with them to the session.} 
\add{We saw that with less than 10 responses total, users typically can read all responses with little support. Similarly, in the case studies we rarely saw participants express interest in much more than 100 responses, at least all at once. 
However, we did not test many tasks that focused on outlier detection, like in a business context where preventing harmful or strange outputs from occurring is more important. It may be that there are some outlier detection tasks where many 100s of responses, even 1000s, are necessary. However, it does seem like once we get to 1000s of responses users typically engage in formal annotation studies. As mentioned in \autoref{sec:annotation}, incorporating formal annotation features into a response inspector could be beneficial.}

\begin{acks}

This material is based on work that is partially funded by an unrestricted gift from Google. 

\end{acks}

\bibliographystyle{ACM-Reference-Format}
\bibliography{main}

\appendix

\section{Features Implemented}

\cut{
\subsection{Implementation Details}
\label{app:implementation}

The renderings we design import responses as a list of JSON objects and render them. ChainForge, and the rendering functionality we built atop it, is written in Typescript.}

\subsection{Exact Matches}
\label{app:exactmatches}

Here we describe in more detail the exact algorithm used to detect exact matches:

We first look for the longest common substring between all possible pairings of responses. We split any substrings that cross sentence boundaries, as this often resulted in exact matches that didn’t represent meaningful text segments. We remove substrings with fewer than three words, as we want to prioritize text segments rather than single words, though this is a variable that could be tuned. All substring matches are then matched to each other across all responses, not just the initial pairing they were originally derived from, such that for each substring we know how many responses in the whole collection it occurs in. Substrings are sorted based on a weighted function of a) how many responses that substring occurs in and b) how long the substring is. For the study, we give length of substring a weight of .75 and number of responses it occurs in a weight of 1.\footnote{Future work could look at how to balance or side-step the tension this algorithm creates between identifying shorter matching text segments that are prevalent across many LLM outputs and longer matching text segments that only occur in a smaller subset of LLM outputs.} Finally, the top k substrings are returned, where k is the minimum of 12 and half the number of all responses in the collection. The value 12 is selected as a maximum value based on the fact that we intended to visualize each match in a different color, and we had a palette of 12 colors which we thought to be visually distinct.

\subsection{Positional Diction Clustering (PDC)}
\label{app:pdc}


\subsubsection{Definition of content similarity.} Our score for content similarity is $(|X in Y| + |Y in X|) / (|X| + |Y|)$, where $X$ and $Y$ are the two sentences, and $|X in Y|$ is the count of words in $X$ that appear in $Y$. This is the same as Bray–Curtis Similarity except that our numerator is the sum of counts rather than two times the min of counts.

\subsubsection{Details of clustering.}
We use a form of single-linkage agglomerative clustering to form groups of sentences across all responses.
First, we split all responses into individual sentences and calculate similarity as described in Section~\ref{alg-grouping}.
We start with each sentence in its own group.
Then, we iterate through all pairs of sentences, sorted by their text similarity, and if a metric that combines text similarity and position similarity is higher than a threshold, we consider combining their groups.
The metric is a linear combination of the two similarities, with a weight of 1.5 on text similarity and a weight of 1 on position similarity, and the threshold is 1.2.
When considering combining groups, we only merge them if at least 70\% of the sentences are from distinct responses.
The groups are ordered by their median normalized location in responses, with ties broken to put sentences from longer documents first.

\subsubsection{Notes on choices}

\begin{itemize}
    \item \textit{Measuring content similarity with exact diction overlap.}
    This  metric is chosen because it maximises the graying out in the second rendering.
    That, in turn, helps users notice what is similar and different across sentences within the same group.
    \item \textit{Using both content and location similarity in the grouping algorithm.}
    This ensures that the groups correspond to parts of the emergent templates in responses.
    Location similarity alone would lead to groups with no coherent meaning.
    Content similarity alone could lead to groups with sentences from the start of one response and the end of another.
    \footnote{For such a group, it would be unclear where to show it in the second rendering. For instance, the mean location, the middle, would not be faithful to the source location for any of the sentences.
    The median location would either be the start or end, which would not be faithful to half of the sentences.}
    \item \textit{Only merging groups if the new group has sentences from a range of responses.}
    This helps form groups that represent emergent patterns across responses, rather than patterns within individual responses.
\end{itemize}

\section{User Study 1}

\subsection{Procedure Details}
\label{app:userstudy1procedure}

Participants were walked through informed consent, and then audio and screen recording began. Participants then filled out a short demographic survey, including questions about their exposure to LLMs. Participants then went through two different tasks---the email rewriting task and the model comparison task---doing each task twice, once in each condition. For the first task, before each interface condition, participants were shown a short tutorial introducing the interface's features.

Within a given task, e.g., email rewriting, the order of the scenarios remained the same and \textit{the interface conditions were counterbalanced}. After completing the task in each scenario, participants filled out a short survey, which included close-ended questions about cognitive load, how realistic the task was, and how well they believed they had performed on the task. After each task, the facilitator conducted a short semi-structured interview with open-ended questions about the utility of the features in both interfaces.
At the end of the study, the facilitator stopped the recording, allowed participants to ask questions about the study, and conferred study payment.
See Appendix \autoref{app:surveys} for survey and interview questions.

\subsection{Determining Time on Task and Time in Feature}

We used the study video recordings to manually determine the amount of time each participant required to finish each task, as well as to identify if participants manually edited their email or used copy/paste in task one, and note if participants used the keyboard search function (command+F) when evaluating model differences in task two. We also used the video recordings to manually log the total time each participant spent in each of the five system features while completing task one and task two. Members of the research team evaluated the timing data for accuracy.

\subsection{Participant Details}
\label{app:participantdata}

\autoref{fig:pre-study} shows participant responses to the pre-study survey questions.

\begin{figure}
    \centering
    
     \begin{minipage}{\textwidth}
        \begin{minipage}[c]{0.3\textwidth}
            \includegraphics[width=\textwidth]{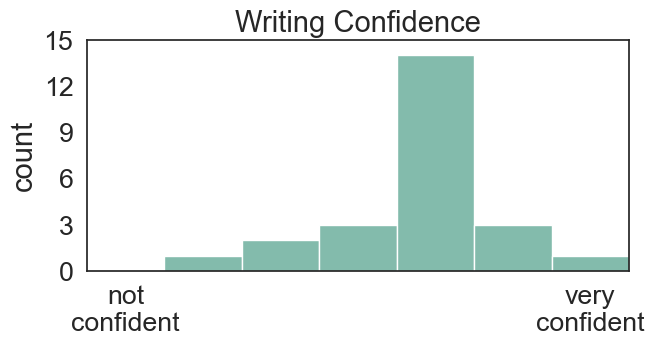}
        \end{minipage}%
        \hspace{0.5em}
        \begin{minipage}[c]{0.4\textwidth}
            (a) ``How confident do you feel when writing important emails? For example, asking for an extension on a project.''
        \end{minipage}
    \end{minipage}

    \vspace{0.5em}

    \begin{minipage}{\textwidth}
        \begin{minipage}[c]{0.3\textwidth}
            \includegraphics[width=\textwidth]{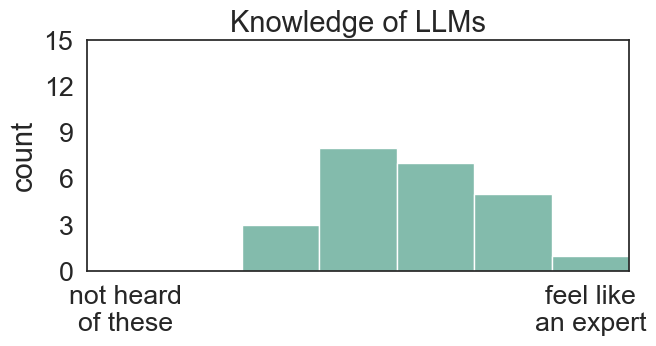}
        \end{minipage}%
        \hspace{0.5em}
        \begin{minipage}[c]{0.4\textwidth}
            (b) ``How much do you know about large language models or chatbots like ChatGPT?''
        \end{minipage}
    \end{minipage}

    \vspace{0.5em}

    \begin{minipage}{\textwidth}
        \begin{minipage}[c]{0.3\textwidth}
            \includegraphics[width=\textwidth]{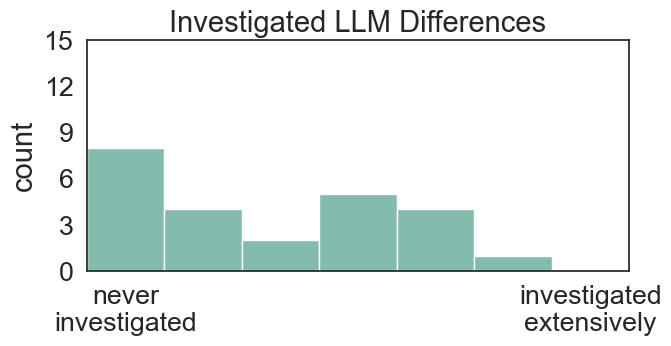}
        \end{minipage}%
        \hspace{0.5em}
        \begin{minipage}[c]{0.4\textwidth}
            (c) ``How much have you investigated the differences between different language models, for instance the differences between GPT-3.5 and GPT-4?''
        \end{minipage}
    \end{minipage}

    \vspace{0.5em}

    \begin{minipage}{\textwidth}
        \begin{minipage}[c]{0.3\textwidth}
            \includegraphics[width=\textwidth]{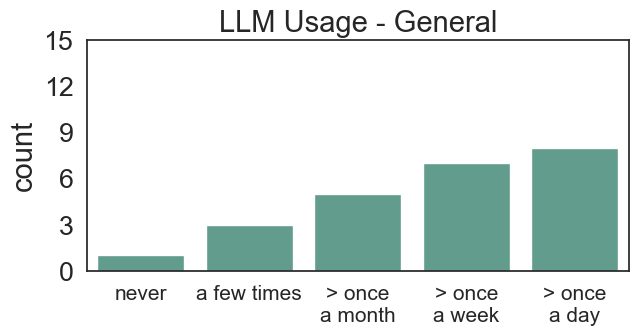}
        \end{minipage}%
        \hspace{0.5em}
        \begin{minipage}[c]{0.4\textwidth}
            (d) ``How often do you use large language models or chatbots like ChatGPT?''
        \end{minipage}
    \end{minipage}

    \vspace{0.5em}

    \begin{minipage}{\textwidth}
        \begin{minipage}[c]{0.3\textwidth}
            \includegraphics[width=\textwidth]{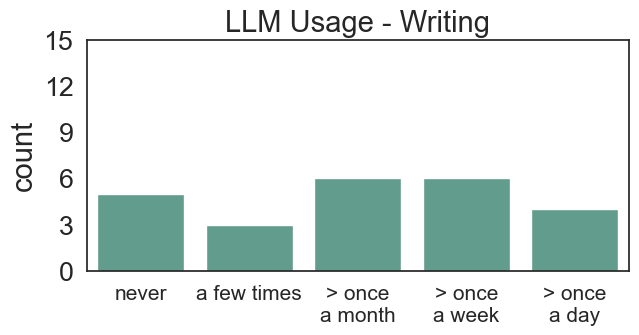}
        \end{minipage}%
        \hspace{0.5em}
        \begin{minipage}[c]{0.4\textwidth}
            (e) ``How often do you use large language models or chatbots like ChatGPT \textbf{for writing}? (That is, not for coding or searching for information.)''
        \end{minipage}
    \end{minipage}
    
    \caption{Participant responses to pre-study questions about experience with writing and LLMs.}
    \label{fig:pre-study}
    \Description{Five histograms of Likert-scale questions in the pre-study survey. First histogram is titled 'Writing Confidence' and the X axis ranges from 'not confident' to 'very confident. It has large peak just right of center. Second histogram is titled 'Knowledge of LLMs' and the X axis ranges from 'not heard of these' to 'feel like an expert'. It has a Gaussian distribution peaked slightly right of center. Third histogram is titled 'Investigated LLM Differences' and the X axis ranges from 'never investigated' to 'investigated extensively'. The distribution is fairly flat but skewed to the left. Fourth histogram is titled 'LLM Usage - General' and the X axis goes from 'never' to 'more than once a day'. Distribution grows decently linearly left to right. Fifth histogram is titled 'LLM Usage - Writing' and the X axis goes from 'never' to 'more than once a day'. Distribution is flat.}
\end{figure}

\subsection{Baseline Interface}
\label{app:baselineinterface}
\autoref{fig:linear} shows the baseline interface using the same example \cut{(writing a short story for a child about a creature)} found in \autoref{sec:feature-descriptions}.

\begin{figure}
\centering
\includegraphics[width=.7\textwidth]{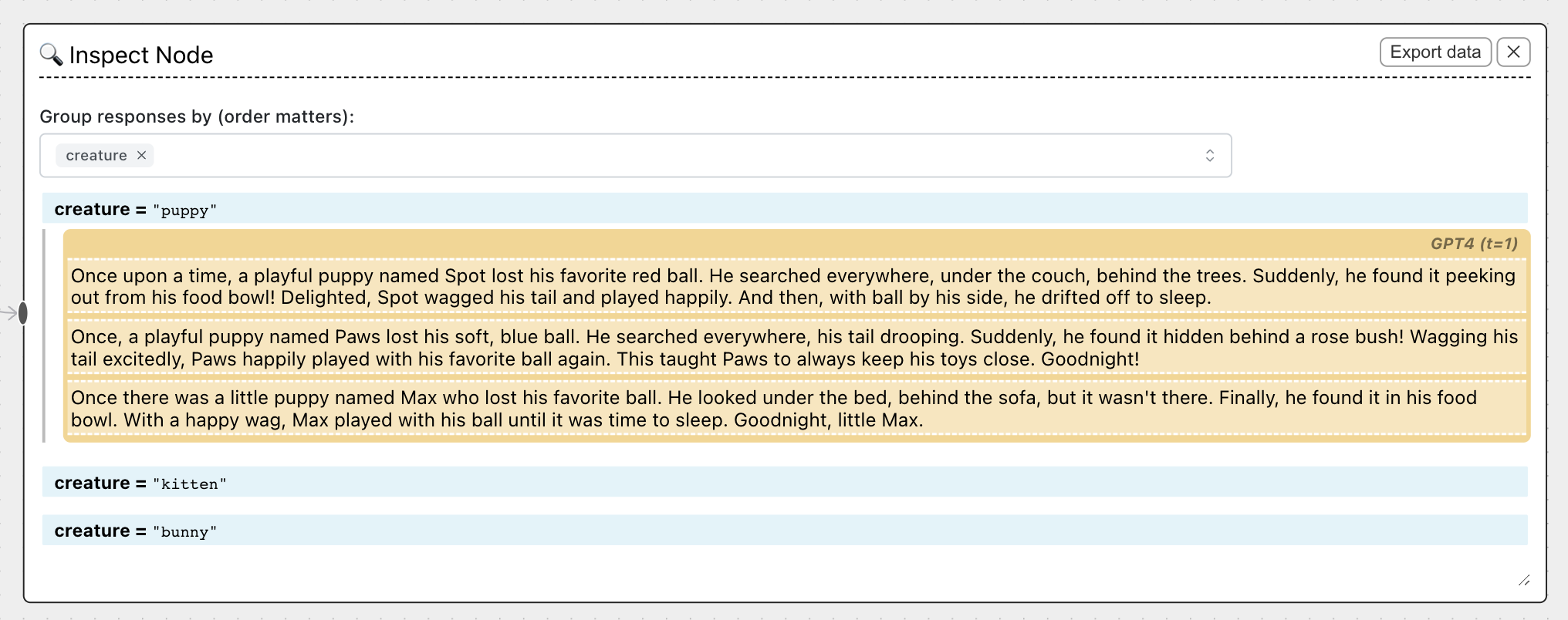}
\caption{The baseline interface allows users to collapse or `hide' groups of responses, such that all or just some groups of responses can be seen at once. Responses can be grouped by model or prompt variation. The example here uses the same example from \autoref{fig:teaser}: ``Write a short story for a five year old child about a \{creature\} that loses something and then finds it again."}
\label{fig:linear}
\Description{Screenshot of the baseline interface. There are three headings: creature=puppy, creature=kitten, and creature=bunny. Under the creature=puppy heading, three responses from the model are listed linear top to bottom. The other headings are `collapsed' with no text below them.}
\end{figure}

\subsection{Task Prompts}
\label{app:prompts}

\subsubsection{Email Re-writing}
    \begin{itemize}
        \item \textbf{Task 1A}: 
        
        {\footnotesize\fontfamily{qcr}\selectfont
        I'm writing an email to my professor asking for a letter of recommendation. Could you rewrite it to be more {style}? Make sure it's less than 100 words.\\
    
        Hi Prof. Sandy,
    
        I took your intro to algorithms class last fall and got an A. I really liked the class and thought you were a great teacher. The class was super hard; can’t believe I got an A, haha! I am applying to a summer internship as a software engineer and was wondering if I could put you down as a reference. If you can’t do it, no worries! Thanks,
    
        [your name]
        } \\

        \item \textbf{Task 1B}: 

        {\footnotesize\fontfamily{qcr}\selectfont
        I’m writing an email to the person who will be my manager for my summer internship. I already have the job offer, but I want to ask to start two weeks later than suggested. Can you rewrite it to be more {style}? Make sure it’s less than 100 words.\\

        Hi Sandra,

        I'm really excited about my summer internship with your team. I am already thinking of project ideas! I know that you said the internship start date for all interns is May 1st. But could I start two weeks later than that? Obviously then I would end two weeks later as well. Totally get it if this isn't possible, but it would really help me out to make this change. Thanks,

        [your name]
        } 
    \end{itemize}

\subsubsection{Model Comparison}
        \begin{itemize}
            \item \textbf{Task 2A}:

            {\footnotesize\fontfamily{qcr}\selectfont
            There's a general audience nonfiction book I'd like to read about neuroscience. I won't have time to read it all. What are some ways I can read just part of it, or skim the book, to get the most out of the book and my time? Keep your response to less than 100 words.} \\

            \item \textbf{Task 2B:}

            {\footnotesize\fontfamily{qcr}\selectfont
            I have a week until an important final exam. I need to study a lot. What do you recommend I do to make sure I perform at my best? Keep your response to less than 100 words.}

        \end{itemize}

\subsection{Pre-Study Survey}
\label{app:surveys}
\small{
\begin{enumerate}
    \item  What is your participant ID?
    (Open-ended response, ID given to participant by researcher)
    \item What is your age 
        \begin{itemize}[label=\textopenbullet]
            \item 18-24
            \item 25-34
            \item 35-44
            \item 45-54
            \item 54+
        \end{itemize}
    \item What is your gender?
        \begin{itemize}[label=\textopenbullet]
            \item Woman
            \item Man
            \item Non-binary
            \item Prefer not to disclose
            \item Other:
        \end{itemize}
    
    \item What kind of student are you?
        \begin{itemize}[label=\textopenbullet]
            \item Undergraduate 
            \item Masters
            \item PhD
            \item Other: 
        \end{itemize}

    \item What is your field of study?
    (Open-ended response)

    \item How confident do you feel when writing important emails? For example, asking for an extension on a project. 
    (Answered on a seven point Likert scale from "Not very confident" to "Very confident")

    \item How much do you know about large language models or chatbots like ChatGPT?
    (Answered on a seven point Likert scale from "I have not heard of these things" to "I feel like I am an expert on these models")

    \item How much have you investigated the differences between different language models, for instance the differences between GPT-3.5 and GPT-4? 
    (Answered on a seven point Likert scale from "I have never investigated this" to "I have investigated the differences extensively")

    \item How often do you use large language models or chatbots like ChatGPT?
         \begin{itemize}[label=\textopenbullet]
            \item Never
            \item I've used them a few times, but not regularly
            \item A few times a month
            \item A few times a week
            \item Once a day or more
        \end{itemize}

    \item How often do you use large language models or chatbots like ChatGPT for writing? (That is, not for coding or searching for information.)
        \begin{itemize}[label=\textopenbullet]
            \item Never
            \item I've used them a few times, but not regularly
            \item A few times a month
            \item A few times a week
            \item Once a day or more
        \end{itemize}
         
\end{enumerate}
}

\subsection{Post-Task Interview}
After completing the second version of each task, participants received the following questions in a post-task interview: 
\small{
\begin{enumerate}
    \item What was your approach toward doing this task?
        \begin{itemize}[label=\textopenbullet]
            \item Was it different in different interfaces (linear v. grid)?
        \end{itemize}
    \item Did the linear or grid view make it easier to find the best output? (Why?)
    \item Was the linear or grid view more overwhelming? (Why?)
    \item Were the grid highlighting or grouping features useful? (Why or why not?)
    \item Is there anything that I didn’t ask about that you want to share?
\end{enumerate}

\subsection{Email Rewriting Survey}
\begin{enumerate}
    \item  What is your participant ID?
    (Open-ended response, ID given to participant by researcher)
    \item Which email rewriting task did you just do?
        \begin{itemize}[label=\textopenbullet]
            \item Asking for a reference letter
            \item Requesting a later start date
        \end{itemize}
    \item Which interface did you have?
        \begin{itemize}[label=\textopenbullet]
            \item Linear inspect node
            \item Grid inspect node
        \end{itemize}
    \item How mentally demanding was the task?
    \\
    (Answered on a seven point Likert scale from "Very low" to "Very high")
    \item How hurried or rushed was the pace of the task?
    \\
    (Answered on a seven point Likert scale from "Very low" to "Very high")
    \item How mentally demanding was the task?
    \\
    (Answered on a seven point Likert scale from "Very low" to "Very high")
    \item How successful were you in accomplishing what you were asked to do?
    \\
    (Answered on a seven point Likert scale from "Very low" to "Very high")
    \item How hard did you have to work to accomplish your level of performance?
    \\
     (Answered on a seven point Likert scale from "Very low" to "Very high")
     \item How insecure, discouraged, irritated, stressed, and annoyed were you?
     \\
     (Answered on a seven point Likert scale from "Very low" to "Very high")
     \item How realistic was the email writing task?
     \\
     (Answered on a seven point Likert scale from "I have never had to write an email for that purpose" to "I have written emails for that purpose in the past")
     \item How close was your \textbf{selected response} to something you would send?
     \\
     (Answered on a seven point Likert scale from "I would never send that response" to "I would definitely send that response")
     \item How close was your \textbf{edited response} to something you would send?
     \\
    (Answered on a seven point Likert scale from "I would never send that response" to "I would definitely send that response")
    
    \textbf{After completing the second task, participants completed this Email Re-writing Survey again and received two additional questions:}

    \item Which interface made the task \textbf{easier}?
    (Answered on a seven point Likert scale from "linear inspect node" to "grid inspect node")
    \item Which interface felt more overwhelming?
    (Answered on a seven point Likert scale from "linear inspect node" to "grid inspect node")

\end{enumerate}
}

\small{
\subsection{Model Comparison Survey}
\begin{enumerate}
    \item  What is your participant ID?
    (Open-ended response, ID given to participant by researcher)
    \item Which model comparison task did you just do?
        \begin{itemize}[label=\textopenbullet]
            \item Advice on how to skim read a book
            \item Advice on how to prepare for finals
        \end{itemize}
    \item Which interface did you have?
        \begin{itemize}[label=\textopenbullet]
            \item Linear inspect node
            \item Grid inspect node
        \end{itemize}
    \item How realistic was the advice topic?
    \\
    (Answered on a seven point Likert scale from "I have never asked for similar advice before" to "I have asked for similar advice before")
    
    \item How mentally demanding was the task?
    \\
    (Answered on a seven point Likert scale from "Very low" to "Very high")
    \item How hurried or rushed was the pace of the task?
    \\
    (Answered on a seven point Likert scale from "Very low" to "Very high")
    \item How successful were you in accomplishing what you were asked to do?
    \\
    (Answered on a seven point Likert scale from "Very low" to "Very high")
    \item How hard did you have to work to accomplish your level of performance?
    \\
    (Answered on a seven point Likert scale from "Very low" to "Very high")
    \item How insecure, discouraged, irritated, stressed, and annoyed were you?
    \\
    (Answered on a seven point Likert scale from "Very low" to "Very high")
    \item Thinking about the \textbf{differences between models}, how well were you able to determine the differences?
    \\
    (Answered on a seven point Likert scale from "I wasn't able to determine many differences" to "I was able to determine most or all of the differences")
    \item Which features were most helpful to detecting \textbf{model differences}?
        \begin{itemize}
            \item[$\Box$] ******I had the linear inspect node*******
            \item[$\Box$] Grid layout
            \item[$\Box$] Exact Matches
            \item[$\Box$] Unique Words
            \item[$\Box$] Similar Sentences
            \item[$\Box$] Groupings View 
        \end{itemize}
        
    \textbf{After completing the second task, participants completed this Model Comparison Survey again and received two additional questions: }

    \item Which interface made the task easier?
    \\
    (Answered on a seven point Likert scale from "linear inspect node" to "grid inspect node")
    
    \item Which interface felt more overwhelming?
    \\
    (Answered on a seven point Likert scale from "linear inspect node" to "grid inspect node")
\end{enumerate}
}

\normalsize

\subsection{Email Rewriting Task Results}
\label{app:emailrewritingresultsdetails}

\subsubsection{Learning Effects: Participants Were Faster the Second Time Around}
Participants found both email rewriting tasks to be quite realistic. \cut{, but felt more rushed in task A. For the first scenario all participants rated the task as above a 5 on a 7-point scale (where 7 is “I have written an email like that before”). For the second scenario (asking for a later start date) 22 out of 24 participants rated the task as above a 5. First we checked if there were significant differences between the first and second email rewriting scenario.} Analyzing their responses to 7-point Likert scale survey questions with a two-tailed Mann-Whitney U test, participants felt significantly more ``hurried or rushed" in task A than task B (p < .01). Analyzing their time on task \cut{in the email rewriting tasks} with a two-tailed t-test, split into the ‘select’ portion of the task and the ‘edit’ position of the task, we found that participants also took significantly more time on task A (select portion: p < .05, edit portion: p < .01). This aligns with their survey responses; participants took more time during task A and felt more rushed. 
For this reason, we split our analysis comparing interface conditions between task A and task B.

\begin{figure}
\centering
\includegraphics[width=.4\textwidth]{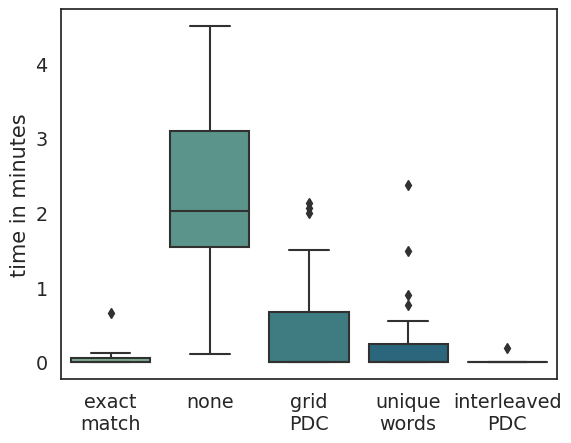}
\caption{ Boxplot of how much time participants spent in each feature when using the exploratory interface. Here `none' refers to being in the grid interface with no highlighting features activated. Note that time spent in a feature does not necessarily indicate that a participant found the feature useful.}
\label{fig:email-featuretiming}
\Description{Boxplot. X axis is the five features, Y axis is time in minutes. All features have a mean of 0 except `none' which has a mean of 2 and a spread from 0 to 4.5.}
\end{figure}

\begin{figure}
     \centering
     \begin{subfigure}[b]{0.4\textwidth}
         \centering
         \includegraphics[width=\textwidth]{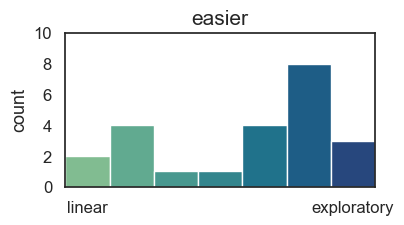}
         \caption{``Which interface made the task easier?''}
         \label{fig:email-easier}
     \end{subfigure}
     \hfill
     \begin{subfigure}[b]{0.4\textwidth}
         \centering
         \includegraphics[width=\textwidth]{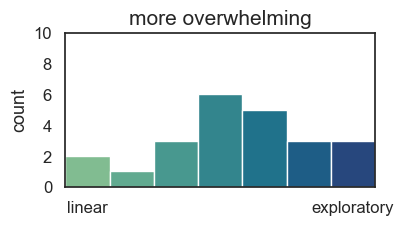}
         \caption{``Which interface was more overwhelming?''}
         \label{fig:email-overwhelm}
     \end{subfigure}
        \caption{Participant interface preferences for the email rewriting task.}
        \label{fig:email-preference}
        \Description{Chart on the left: histogram from `linear' to `exploratory' with seven buckets. Histogram is roughly U-shaped, with slightly more participants preferred exploratory. Chart on the right: also histogram from `linear' to `exploratory' with seven buckets. Shape is more Gaussian, with a peak in the middle. }
\end{figure}

\subsection{Model Differences Task Results}

\begin{figure}
     \centering
         \includegraphics[width=.5\textwidth]{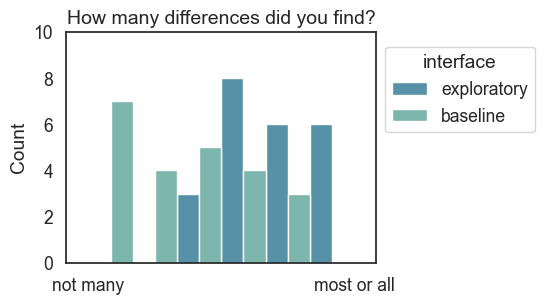}
        \caption{Participant responses to the question ``Thinking about the differences between models, how well were you able to determine the differences?''}
        \label{fig:modeldiff-success}
        \Description{Interleaved histogram. X axis is seven buckets from `not many' to `most or all'. Histogram of exploratory has a peak between 4 and 5. Histogram of baseline has a peak between 3 and 4.}
\end{figure}

\subsection{Interview Guideline for Case Study}
\label{app:casestudyinterviewguideline}
\small{
\begin{itemize}
    \item Explain consent details (recording, how data will be aggregated, anonymized, and used for research purposes, their ability to opt out, and their right to request a copy of the paper) and verbally request for their consent to record. Begin recording via Zoom.
    \item Demo interface
    \item Request for participant to open the interface and share their screen
    \item Send API keys; get them running on the interface
    \item Ask them to think aloud through their process; guide them to more interesting prompts if necessary; ask them what they are thinking about or seeing. Encourage them to try out all the features. Explain how features work when requested.
    \item Sample interview questions:
        \begin{itemize}
            \item What kinds of tasks or information do you want to look at? Compare prompts, models, prompt variables? Look at long tail distribution, or set of standard responses? 
                \item Observations: What feature do they try first? Second? Ask why they tried particular prompts.
            \item Does this tool support your exploration and inspection of outputs?
            \item What works? What is missing?
            \item How would you complete the selected task without this tool?
            \item How does this experience differ from your prior interaction with LLMs?
            \item The goal is to aid in skimming and comparison of responses: does it do that?
        \end{itemize}
\end{itemize}
}

\normalsize
\subsection{Formative Interview Guideline}
\label{app:formativeinterviewguideline}

The interview guideline was meant to be extremely broad, given the range of kinds of LLM-based tasks participants were engaged in. Here we list all questions, however many were not relevant for certain participants, and many custom follow-up questions were asked depending on the context in which the participant was working.

\small{
\begin{itemize}
    \item How do you select models or design prompts?
    \item Do benchmarks work for you? Do you use metrics? Trial and error? To what extent does quantitative evaluation or benchmarks work for your use case?
    \item Do you ever regenerate or compare outputs? What about between prompts or models?
    \item Have you noticed models changing over time? Have you tried different models?
    \item How do you know your system is working? What’s success for you?
    \item How often do you update models / prompts?
    \item What might help qualitative evaluation?
    \item What does qualitative evaluation look like? Are you comparing models, prompts, other things? Do you rely on annotations, having people read outputs and discuss, or other methods?
    \item Does any evaluation involve comparing multiple outputs in a way other than via annotation? Do you ever look for qualitative differences between outputs? e.g. “This prompt produces more verbose and flowery outputs.”
    \item When you have people manually inspect outputs, how many outputs does any one person typically look at at a time?
\end{itemize}
}

\end{document}